\documentclass{aa}  

\usepackage{graphicx}
\usepackage{longtable}
\usepackage{txfonts}
\usepackage[usenames,dvipsnames]{color}

\begin{document} 

\bibliographystyle{aa.bst}
   \title{The different origins of magnetic fields and activity in the Hertzsprung gap stars, OU~Andromedae and 31~Comae\thanks{Based on observations obtained at the T\'elescope Bernard Lyot (TBL) at Observatoire du Pic du Midi, CNRS/INSU and Universit\'e de Toulouse, France.}}

   \author{A. Borisova
          \inst{1},
          M. Auri\`{e}re \inst{2,3},
          P. Petit \inst{2,3}, 
          R. Konstantinova-Antova
          \inst{1,2},          
          C. Charbonnel \inst{4,3}
          \and
          N. A. Drake \inst{5,6}
          }
\institute{Institute of Astronomy, Bulgarian Academy of Sciences,
              71 Tsarigradsko Shosse blvd, 1784 Sofia, Bulgaria \\
             \email{aborisova@astro.bas.bg}
             \and
             Universit\'{e} de Toulouse, UPS-OMP, Institut de Recherche en Astrophysique et Plan\'{e}tologie, Toulouse, France
             \and
             CNRS, UMR 5277, Institut de Recherche en Astrophysique et Plan\'etologie,
             14 av. Edouard Belin, 31400 Toulouse, France          
             \and
          Department of Astronomy, University of Geneva,
          Chemin des Maillettes 51, 1290 Versoix, Switzerland
            \and
           Saint Petersburg State University, Saint Petersburg,
           Universitetski pr. 28, 198504 Saint Petersburg, Russia 
             \and        
           Observat\'orio Nacional/MCTI, Rua General Jos\'e
           Cristino 77, 20921-400 Rio de Janeiro, Brazil           }

   \date{Received May 31, 2015; accepted Jun 22, 2015}

% \abstract{}{}{}{}{} 
% 5 {} token are mandatory
 
  \abstract
   {When crossing the Hertzsprung gap, intermediate-mass stars develop a convective envelope. Fast rotators on the main sequence, or Ap star descendants, are expected to become magnetic active subgiants during this evolutionary phase.}
   {We compare the surface magnetic fields and activity indicators of two active, fast rotating red giants with similar masses and 
   spectral class but different rotation rates - OU~And ($P_{rot}$=24.2~d) and 31~Com ($P_{rot}$=6.8~d) -  to address the question of the origin of their magnetism and high activity.}
   {Observations were carried out with the Narval spectropolarimeter in 2008 and 2013. We used  the least squares deconvolution (LSD)
   technique to extract Stokes~V and I profiles with high  S/N   to detect Zeeman signatures of the magnetic field of the stars. We then  provide Zeeman-Doppler imaging (ZDI), activity indicators monitoring, and a precise estimation of stellar parameters. We use state-of-the-art stellar evolutionary models, including rotation, to infer the evolutionary status of our giants, as well as their initial rotation velocity on the main sequence, and we interpret our observational results in the light of the theoretical Rossby numbers.}
   {The detected magnetic field of OU\ Andromedae (OU\ And) is a strong one. Its longitudinal component $B_l$ reaches 40~G and presents an about sinusoidal variation with reversal of the polarity. The magnetic topology of OU~And is dominated by large-scale elements and is mainly poloidal with an important dipole component, as well as a significant toroidal component. The detected magnetic field of 31~Comae (31~Com) is weaker, with a magnetic map showing a more complex field geometry, and poloidal and toroidal components of equal contributions. The evolutionary models show that the progenitors of OU~And and 31~Com must have been rotating at velocities that correspond to 30 and 53~$\%$,
respectively, of their critical rotation velocity on the zero age main sequence. Both OU~And and 31~Com have very similar masses (2.7 and 2.85~M$_{\odot}$, respectively), and they both lie in the Hertzsprung gap.} %at the beginningof the so-called magnetic strip.}}
    %CC20150403 OU~And is the sprunt of a moderate rotator and 31~Com \CC{relatively fast rotator} should have rotated near the critical velocity on the main sequence.}
   {OU~And appears to be the probable descendant of a magnetic Ap~star, and 31~Com  the descendant of a relatively fast rotator on the main sequence. Because of the relatively fast rotation in the Hertzsprung gap and the onset of the development of a convective envelope, OU~And also has  a dynamo in operation.}

   \keywords{ stars: individual: OU~And - stars: individual: 31~Com - stars: magnetic field -stars: late-type}
   \titlerunning{Magnetic fields and activity in OU~And and 31~Com}
   \authorrunning{A. Borisova et al. }
  \maketitle

\section{Introduction}
  \label{section:Intro}
Recent studies report the detection of magnetic fields with strengths that range from a few tenths to a hundred gauss at the surface of single red giants of intermediate mass \citep{renadaiaus14,big_paper}. 
This raises the question of the origin and evolution of magnetic fields during the stellar evolution. An interesting evolutionary stage is the subgiant phase. 
When intermediate mass stars leave the main sequence and cross the Hertzsprung gap, they develop a convective envelope, and their surface rotation rate decreases as a consequence of the increase of their radius.
In ordinary giant stars, $\alpha-\omega$ dynamo is then favored during the so-called first dredge-up at the base of the red giant branch, owing to structural changes of the convective envelope that imply an increase of the convective turnover timescale and a temporary decrease of the Rossby number \citep[so-called magnetic strip, see][]{charbonnel_2016,big_paper} ; in parallel at that phase, the large-scale fossil magnetic field of Ap star descendants has not yet been  strongly diluted. A number of Hertzsprung gap-crossers are detected in X-rays as active stars \citep{gondoin_99, gondoin_2005, ayres_98, ayres_2007}. For these stars, magnetic activity is a result of a remnant of relatively fast rotation, because the convective envelope is not yet well developed.

 An essential point is to distinguish between descendants of stars hosting strong fossil magnetic fields on the surface of the main sequence (magnetic Ap stars) and the ones generating their magnetic field as a result of their evolution during the Hertzsprung gap. 
In this paper we present a study of the magnetism and activity of two subgiant stars, OU~And and 31~Com, that are very similar in terms of mass, spectral type, activity properties, evolutionary stage (i.e., they both lie in the Hertzsprung gap stepping up to the edge of the so-called magnetic strip), and strong X-ray activity, but which have significantly different rotational rates.

\begin{table*}[htb]
\caption{Fundamental parameters for OU~And and 31~Com}
\label{tab:stars}
\centering
\begin{tabular}{lcccccc}

\hline 

            &    OU~And    &         & \vline &31~Com &        \\
            & HD 223460              &        & \vline &    HD 111812        & \\
& & & \vline & &   \\
Parameter                & Value     & Reference  & \vline & Value & Reference   \\
\hline
& & & \vline & &   \\
$V$, mag                 & 5.86       & [1]        & \vline & 4.93  & [1] \\
Distance, pc             &$129.7^{-5.5}_{+6.0}$ & [2] & \vline & $89.1^{+1.9}_{-1.7} $ & [2]           \\
Sp Type                  & G1 III    & [3]         & \vline & G0 III& [3]       \\
$T_{\rm eff}$, K         & 5360&[4]  & \vline & $5660 \pm 42$  & [5] \\
$L_{x}$, $10^{27}$ erg/s & 8203      & [6]   & \vline & 6325  & [6]            \\
Radius, $R_{\odot}$      & 9.46 & [4]   & \vline &8.5 & [4]      \\
Mass, $M_{\odot}$        &  2.85      & [4]   & \vline & 2.75       & [4]       \\
Luminosity, $L_{\odot}$  &  71.2     & [4]  & \vline &73.4 & [4]          \\ 
$\log g$                 & 2.8       & [4]   & \vline &2.97 & [4]       \\
$[\rm Fe/H]$             & -0.07     & [4]   & \vline & $0.00 \pm 0.08$ & [9]         \\
$v\sin i$, km\,s$^{-1}$  & 21.5     &   [8]    & \vline & $ 67 \pm 2 $ & [5]           \\
$P_{\rm rot}$, days      & 24.2     & [7]    & \vline & $6.8 \pm 0.006$  & [5]\\
Radial velocity, km\,s$^{-1}$ & -2.47 & [8]  & \vline & $+0.10 \pm 0.33$ & [5]   \\
\hline 
\end{tabular}
\\
\noindent 
\tablefoot{[1] Hipparcos catalogue, ESA 1997,[2]{\cite{hipp}},[3]{\cite{gray_2001}},[4] Present paper, [5]{\cite{strass_2010}},[6]{\cite{gondoin_99}}, [7]{\cite{strass_99}}, [8]{\cite{deMedeiros_99}},[9]{\cite{heiter_2014}}}

\end{table*}

OU~And (HD~223460, HR~9024) is a single, G1~III giant \citep{gray_2001} with moderate emission in \ion{Ca}{ii}~H\&K line \citep{cowley_79}. It is a moderate rotator with $v\sin i\sim21$~km\,s$^{-1}$ for its position in the Herzsprung gap with $T_{\rm eff}\sim5360$~K and $M=2.85~M_{\odot}$ (see Table~\ref{tab:stars} and \S~5.1).
The photometric variability of the star was discovered by \cite{hopkins_85} who determined a photometric period $23.25\pm0.09$. \cite{strass_88} provided a long-term monitoring of the brightness of the star and confirmed that the photometric period and light curve amplitude undergoes smooth changes. They estimated a photometric period of $22.6\pm0.09$. The latest photomeric data of the star are presented by \cite{strass_99} and reveal an uncertain period of $24.2$ days and $V$ amplitude of about $0.01$~mag. Amplitude and period changes,  as in OU~And, are also observed in other chromospherically active giants and may indicate cyclic activity like in the Sun. OU~And is chromospherically active with X-ray luminosity variations. Series of highly ionised iron lines, several Lyman lines of hydrogen-like ions and triplet lines of Helium-like ions in the X-ray region \citep{gondoin_2003, ayres_2007} are observed in its spectra.

31~Com (HD~111812) is a single G0~III, \citep{gray_2001}, rapidly 
rotating giant with $v\sin i\sim67$~km/s, $T_{\rm eff}\sim5660$~K and $M=2.75~M_{\odot}$ \citep[][ see Table~\ref{tab:stars} and \S~5.1]{strass_2010}. The star is a variable with a very low light curve amplitude and rotational modulation with a period of $\sim~6.8$~d. Its light curve appears to be complex, with period and shape changes and multiple peaks 
in the periodogram \citep{strass_2010}. The star shows chromospheric and coronal activity with \ion{Ca}{ii}~H\&K line emission, 
super-rotationally broadened coronal and transition-region lines, and strong X-ray activity \citep{gondoin_2005}.

Table~\ref{tab:stars} summarizes the fundamental parameters for OU~And and 31~Com which we have used in this paper, with their relevant references. The stellar radius is obtained from the Stefan-Boltzmann law using the adopted values for the stellar effective temperature and luminosity.

\section{Observations and data processing}
\label{section:Observ}
Observations were carried out with the Narval spectropolarimeter of the 2~m 
telescope Bernard Lyot at Pic du Midi Observatory, France, during two consecutive semesters of 2013. 31~Com 
was observed in the first semester of  2013 during 10 nights of April and May, while OU~And was 
observed in the second semester of 2013, during 13 nights in September and October. Additionally, we used six OU~And observations obtained with the same instrumental configuration in September 2008. The total exposure time for each observation of OU~And was 40 minutes with the exception of the first two observations : on 14 September 2008 it was 4 minutes and on 16 September 2008 it was 16 minutes.The total exposure time for each observation of 31~Com was $\sim33.3$~minutes. Detailed information for these observations is presented in Tables~\ref{tab:ouand} and \ref{tab:31com} for OU~And and 31~Com, respectively. Observational data obtained in 2013 for OU~And were collected during three rotations of the star, while for 31~Com they were collected in four stellar rotations. In 2013, we  obtained relatively good phase coverage for OU~And, with some small phase gaps near phases 0.6 and 0.9. The observational data set for 31~Com suffers from one large data -gap, of about 40\% of the rotational 
cycle.

Narval is a twin of ESPaDOnS spectropolarimeter \citep{donati_2006b}. This consists of a polarimetric unit connected by optical fibers to a cross--dispersed \'{e}chelle spectrometer. The instrument has high-resolution ($R = \lambda \diagup \Delta\lambda \simeq 65000$) 
and wavelength coverage from about 369 to 1048~nm. In the spectropolarimetric mode two orthogonally 
polarized spectra are recorded in a single exposure over the whole spectral band. 
We obtained standard circular-polarization observations that consist of a series of four sub-exposures, 
between which the half-wave retarders (Fresnel rhombs) are rotated to change paths (and the spectra positions on
the CCD) of the two orthogonally polarized light beams. This way the spurious polarization signatures are reduced.
Typically diagnostic null spectra $N$ are also included in the observational data. In principle, these spectra should not contain polarization signatures and serve to ensure the absence of any spurious signal in the Stokes $V$ spectra.

The data were reduced using the Libre ESpRIT \citep{donati_97} software for automatic spectra extraction. The software includes wavelength calibration, heliocentric frame correction, and continuum normalization. Afterwards, a data reduction extracted spectra are recorded in an ASCII file and they consist of the normalized Stokes~$I$ ($I \diagup I_c$) and Stokes~$V$ ($V \diagup I_c$) intensities and the Stokes~$V$ uncertainty, $\sigma_{v}$ as a function of wavelength, (where $I_c$ represents the continuum level). 

The least-Squares deconvolution method, \citep[here and after - LSD, ][]{donati_97} was applied to all of the observations to perform extraction of the mean Stokes~$V$ and Stokes~$I$ photospheric profiles. This is a multi-line analysis method and assumes that all the spectral lines have the same profile, scaled by a certain factor and that the observed spectrum can be represented as a convolution of single mean line profile with the so-called line mask. In the case of cool stars, this technique  usually averages several thousand spectral lines and thus significantly increases the S/N  in the mean line profile. The line mask was computed by the use of spectral synthesis based on the \cite{Kurucz_18CD} models, and represented theoretical spectrum of unbroadened spectral lines with appropriate wavelength, depth, and Lande factor, \cite{donati_97}. For the present observations, we used a digital mask calculated for an effective temperature of 5750~$K$ and $\log{g}=3.5$ and with about 8 900 spectral lines for 
both stars. We used the atomic transition data provided by the VALD database, \cite{VALD}.
 Owing to the LSD technique the S/N  in the Stokes~$V$ profiles for both stars was increased by about 40 times. This is presented in the last column of Tables~\ref{tab:ouand} and \ref{tab:31com}. 

\begin{table*}[htb]
\caption{Spectropolarimetric observations, activity indicators, and $B_l$ measurements for OU~And in 2008 and 2013 (see \S~\ref{section:MFandAI}).}
\label{tab:ouand}
\centering
\begin{tabular}{lccccccccc}

\hline 
 & & & & \vline &       &   & & &  \\
Date         & HJD         & Phases & $S \diagup N$ & \vline & S-index&H$\alpha$     & \ion{Ca IRT} & B$_l$   & $\sigma$ \\
             & 2450 000 +  &        &               & \vline &          &index         & index  & (G)      & (G)    \\
\hline
 & & & & \vline &       &   & & &  \\
14 Sep. 2008 & 4724.46     & 0.74   &11566 & \vline & 0.492 & 0.3338  &0.873 & -28.0  & 5.5  \\
16 Sep. 2008 & 4726.52     & 0.83   &11437 & \vline & 0.469 & 0.3183  &0.856 & -24.8  & 3.2  \\
19 Sep. 2008 & 4729.57     & 0.95   &36567 & \vline & 0.498 & 0.3391  &0.869 & -10.3  & 1.7  \\
21 Sep. 2008 & 4731.41     & 0.03   &23403 & \vline & 0.515 & 0.3499  &0.878 & +5.6   & 2.7  \\
25 Sep. 2008 & 4735.43     & 0.20   &37585 & \vline & 0.517 & 0.3421  &0.879 & +30.0  & 1.7  \\
29 Sep. 2008 & 4739.52     & 0.36   &42200 & \vline & 0.515 & 0.3434  &0.870 & +40.9  & 1.5  \\
02 Sep. 2013 & 6538.57     & 0.71   & 39601& \vline & 0.525 & 0.3366  &0.887 & +36.1  & 1.6   \\
08 Sep. 2013 & 6544.52     & 0.95   & 34881& \vline & 0.483 & 0.3269  &0.864 & +33.0  & 1.8   \\
10 Sep. 2013 & 6546.50     & 0.03   & 39508& \vline & 0.504 & 0.3335  &0.870 & +20.6  & 1.6   \\
15 Sep. 2013 & 6551.54     & 0.24   & 40125& \vline & 0.496 & 0.3422  &0.866 & -24.7  & 1.6   \\
17 Sep. 2013 & 6553.52     & 0.32   & 31367& \vline & 0.521 & 0.3515  &0.887 & -345.0 & 2.1   \\
19 Sep. 2013 & 6555.51     & 0.41   & 38155& \vline & 0.494 & 0.3395  &0.874 & -19.5  & 1.7   \\
21 Sep. 2013 & 6557.50     & 0.49   & 40193& \vline & 0.496 & 0.3397  &0.876 & -2.9   & 1.6   \\
23 Sep. 2013 & 6559.49     & 0.57   & 42731& \vline & 0.484 & 0.3322  &0.873 & +17.1  & 1.5   \\
06 Oct. 2013 & 6572.47     & 0.11   & 29819& \vline & 0.473 & 0.3357  &0.868 & -5.0   & 2.1   \\
08 Oct. 2013 & 6574.46     & 0.19   & 40784& \vline & 0.484 & 0.3652  &0.854 & -24.6  & 1.5   \\
11 Oct. 2013 & 6577.57     & 0.32   & 37216& \vline & 0.482 & 0.3421  &0.862 & -27.7  & 1.7   \\
13 Oct. 2013 & 6579.57     & 0.40   & 21087& \vline & 0.485 & 0.3595  &0.866 & -18.6  & 3.0   \\
31 Oct. 2013 & 6597.38     & 0.14   & 29209& \vline & 0.534 & 0.3466  &0.887 & -145.0  &2.2    \\
  
\hline
\end{tabular} 
\\
\tablefoot{For ephemeris computation HJD$_{0}$ = 2454 101.5 and rotational period of 24$^{d}$.2 were used. Presented $S\diagup N$  is for LSD Stokes~$V$ profiles. Scaling parameters (line depth, wavelength, and land\'e factor) are equal to 0.530, 562~nm, and 1.29, respectively.}
\end{table*}

\section{Magnetic field detection and strength. Activity indicators behavior.}
\label{section:MFandAI}
The output of the LSD procedure contains the mean Stokes~$V$ and $I$, the diagnostic null ($N$) profile, as well as a statistical test for the detection of Zeeman signatures in Stokes~$V$ profiles. The test is performed inside and outside spectral lines and estimates the magnetic field detection probabilities, as described by \cite{donati_92, donati_97}. The "definite detection" is assumed if the probability for signal detection in the spectral line is greater than 99.999\%, "marginal detection" -- if it falls in the range 99.9\% and 99.999\%, and no detection otherwise. We examine the signal in the $N$ profile and also outside the spectral line and for a reliable detection the one without a signal in the latter two  is accepted. Mean surface-averaged longitudinal magnetic field ($B_l$, G) was computed by the use of the first order moment method \citep{rees_79}, adapted to LSD profiles \citep{donati_97,wade_2000}. 

The activity of the stars were also examined using  traditional spectral activity indicators, \ion{Ca}{ii}~H\&K, S-index, H$\alpha,$ and \ion{Ca}~IR triplet indexes, \cite{marsden_2014}. We calculated the S-index as it is defined for the Mount Wilson survey, \citep{duncan_91} and our computational procedure was calibrated for Narval spectra, see \cite{big_paper}. The H$\alpha$ and \ion{Ca}{ii}~IR triplet indexes are introduced by \cite{gizis_2002} for the Palomar Nearby Spectroscopic Survey and by \cite{petit_2013} for the Betelgeuse cool supergiant, respectively. In this study, we used a slightly modified H$\alpha$ index with the continuum taken slightly farther from the line center to avoid the influence of broadening that is due to the active chromosphere and fast rotation.

\begin{figure*}
\centering
\includegraphics[width=0.95\columnwidth]{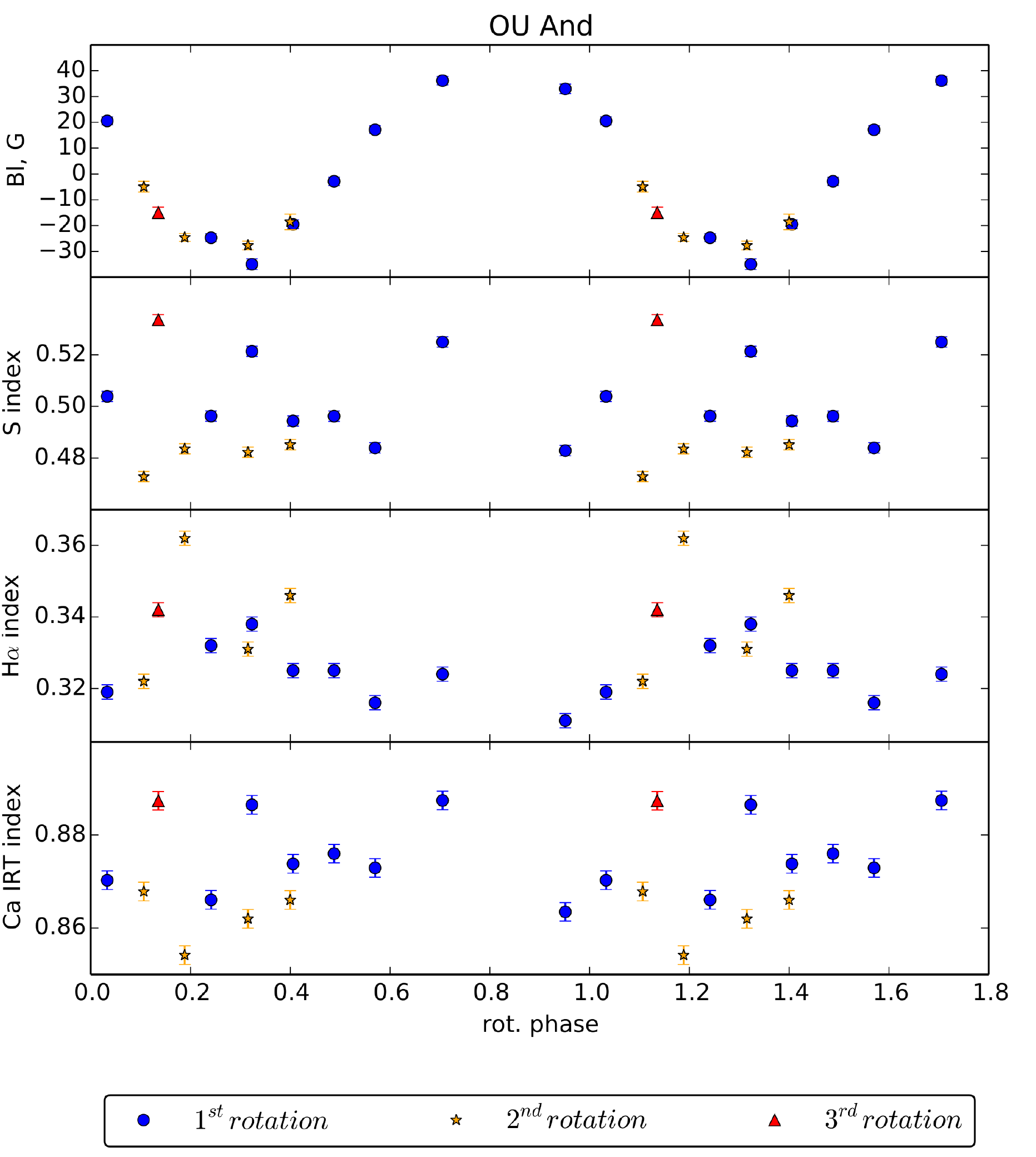}
\includegraphics[width=0.95\columnwidth]{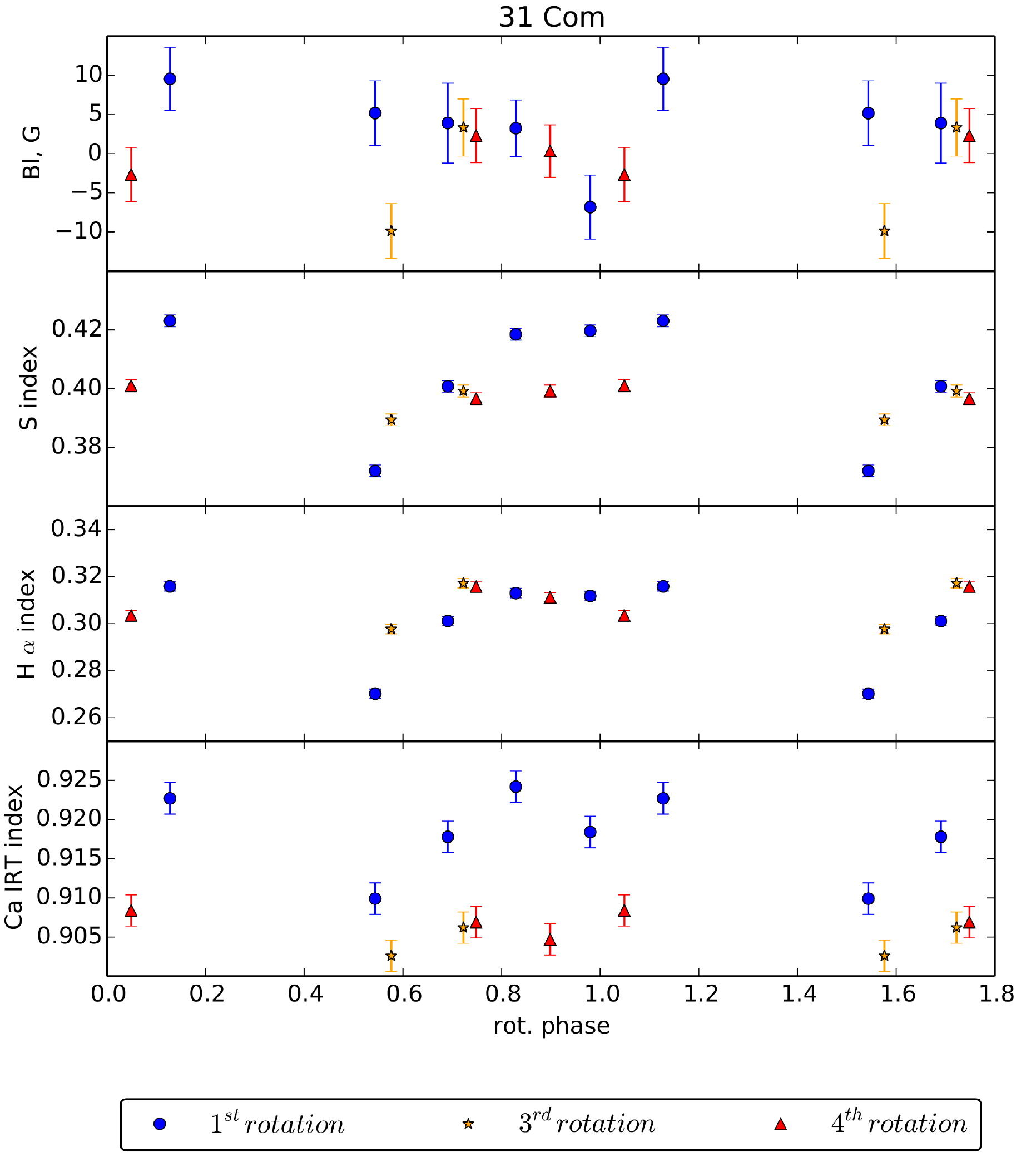}
\caption{$B_l$, S-index, H$\alpha$, and \ion{Ca}{ii}~IR triplet indexes behavior for OU~And (left) and 31~Com (right). For both stars, phases are computed for ephemeris HJD$_{0}$ = 2454 101.5 and the rotational period of 24.2~d for OU~And and 6.8~d for 31~Com.}
\label{fig:bl_act_rot}
\end{figure*}

Observational results for detected magnetic field and for the activity indicators are presented in Tables~\ref{tab:ouand} and \ref{tab:31com}, as well as in Fig.~\ref{fig:bl_act_rot}. The tables columns are: Date of observation, HJD, rotational phases, $S \diagup N$  in LSD Stokes~$V$ profile, S-index, H$\alpha$ index, \ion{Ca}{ii}~IR index, $B_l$, and the corresponding accuracy in G. Rotational phases are computed for ephemeris HJD$_{0}$ = 2454 101.5 and rotational period of 24.2~d for OU~And and 6.8~d for 31~Com. Preferred choice for the rotational period of OU~And is discussed in Section~\ref{subsection:MMOUAnd}.
 
\begin{figure}
\centering

\includegraphics[width=1.0\columnwidth]{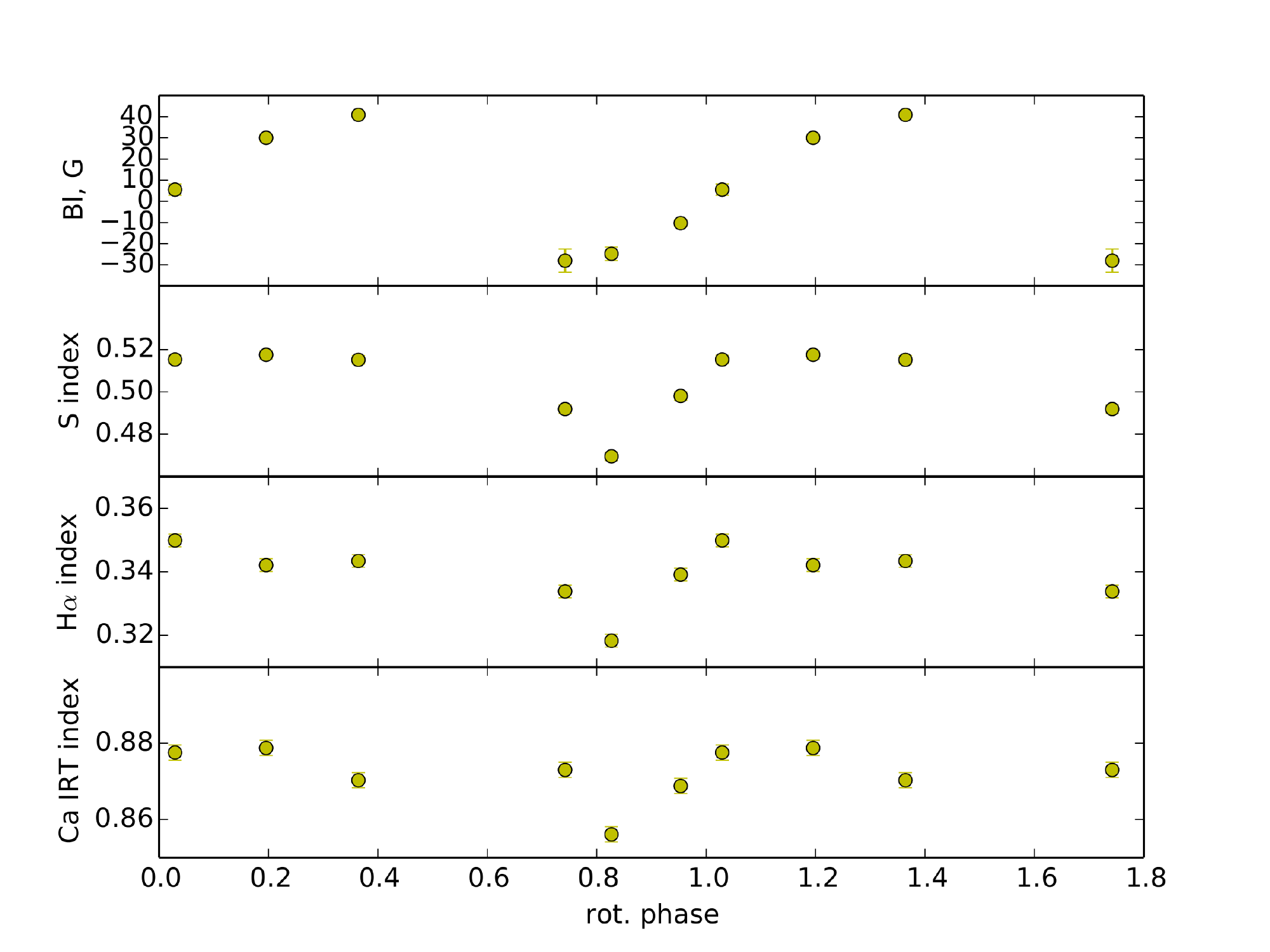}

\caption{B$_l$, S-index, H$\alpha$, \ion{Ca}{ii}~IR triplet indexes behavior for observations of OU~And gathered in 2008. The plots follow the same definitions as in Fig.\ref{fig:bl_act_rot}.}

\label{fig:ouand08_b}
\end{figure}

Figure \ref{fig:bl_act_rot} represents $B_l$, S-index, H$\alpha$, and \ion{Ca}{ii}~IR triplet indexes for OU~And (in 2013) on the left panel and for 31~Com on the right panel. Activity indicators and $B_l$ plots for OU~And in 2008 are presented in Fig. \ref{fig:ouand08_b}. Rotational phases are computed the same way as for Fig.~\ref{fig:bl_act_rot}. For OU~And, the error bars on the plots are of about the same size as the plot symbol and they are not visible.
We have definite magnetic field detection for all the spectra, except for one of 31~Com, which was obtained on the 20 April 2013 with a marginal detection. OU~And data reveal a strong Zeeman signature in the Stokes~$V$ profile with a structure that has several peaks, 
while for 31~Com we have got relatively weaker signatures, with strong noise contribution and a much more complex structure. 

For OU~And, the longitudinal magnetic field, $B_l$, shows a very good rotational modulation and we have well-sampled data at the minimum $B_l$. The maximal values of $B_l$ are not well covered with data. We observe anticorrelation between the $B_l$ and H$\alpha$ index. There is also a good correlation between the two Calcium indexes, which indicates that the measurements really concern the \ion{Ca}{ii} contribution. There is some scatter between the different rotations for the activity indicators, although not for the magnetic field. For most activity proxies, some scatter is observed between successive rotation cycles. The measurements obtained at different stellar rotations agree much better for the longitudinal magnetic field.

For 31~Com we see clear $B_l$ minimum and a smooth variation with rotation. However, because the longitudinal field estimates result from an integration over the full width of the line profile,  they do not therefore capture the full complexity of the Stokes V profile, and hence the magnetic structure. The indexes show a good correlation, but we have a large data gap between phases 0.2 and 0.5. The S-index and \ion{Ca}{ii}~IR triplet index show a decrease in the third rotation. The $B_l$ measurement on 20 April 2013 is excluded because of the marginal magnetic field detection.

\begin{table*}[htb]
\caption{Spectropolarimetric observations, activity indicators and $B_l$ measurements for 31~Com in 2013.}
\label{tab:31com}
\centering
\begin{tabular}{lcccccccccc}

\hline
&  & & & \vline&   &    & &  &  \\
Date         & HJD         & Phases  &$S \diagup N $ &\vline&S-index & H$\alpha$     & \ion{Ca IRT}    & $B_l$   &$ \sigma $  \\
             & 2450 000 +  &         &               &\vline&            & index         & index      & (G)       & (G)      \\
\hline
&  & & & \vline&   &    & &  &  \\
20 Apr. 2013 &  6403.597   & 0.54  & 48735    & \vline&       0.372& 0.2702         &0.910 & 5.2     &  4.1   \\
21 Apr. 2013 &  6404.599   & 0.69  & 39671    & \vline&       0.401& 0.3011         &0.918 & 3.9     &  5.1    \\
22 Apr. 2013 &  6405.537   & 0.83  & 56620    & \vline&       0.419& 0.3130         &0.924 & 3.2     &  3.6    \\
23 Apr. 2013 &  6406.564   & 0.98  & 49814    & \vline&       0.420& 0.3118         &0.918 & -6.8    &  4.1    \\
24 Apr. 2013 &  6407.569   & 0.13  & 50150    & \vline&       0.423& 0.3159         &0.923 & 9.5     &  4.0    \\
04 May. 2013 &  6417.420   & 0.58  & 57643    & \vline&       0.389& 0.2977         &0.903 & -9.9    &  3.5    \\
05 May. 2013 &  6418.414   & 0.72  & 55198    & \vline&       0.399& 0.3172         &0.906 & 3.3     &  3.7    \\
12 May. 2013 &  6425.390   & 0.75  & 59246    & \vline&       0.397& 0.3158         &0.907 & 2.3     &  3.4    \\
13 May. 2013 &  6426.410   & 0.90  & 59632    & \vline&       0.399& 0.3112         &0.905 & 0.3     &  3.4    \\
14 May. 2013 &  6427.433   & 0.05  & 58354    & \vline&       0.401& 0.3035         &0.908 & -2.7    &  3.5    \\ 
   
\hline
\end{tabular}
\\
\tablefoot{For ephemeris computation HJD$_{0}$ = 2454101.5 and a rotational period of 6$^{d}$.8 were used. The presented $S \diagup N$  is for LSD Stokes~$V$ profiles. Scaling parameters (line depth, wavelength, and land\'e factor) are equal to 0.530, 562~nm, and 1.29, respectively.} 

\end{table*}

\section{Magnetic mapping}
\label{section:MMapping}

\begin{figure*}[htb]
\centering
\includegraphics[width=0.80\columnwidth]{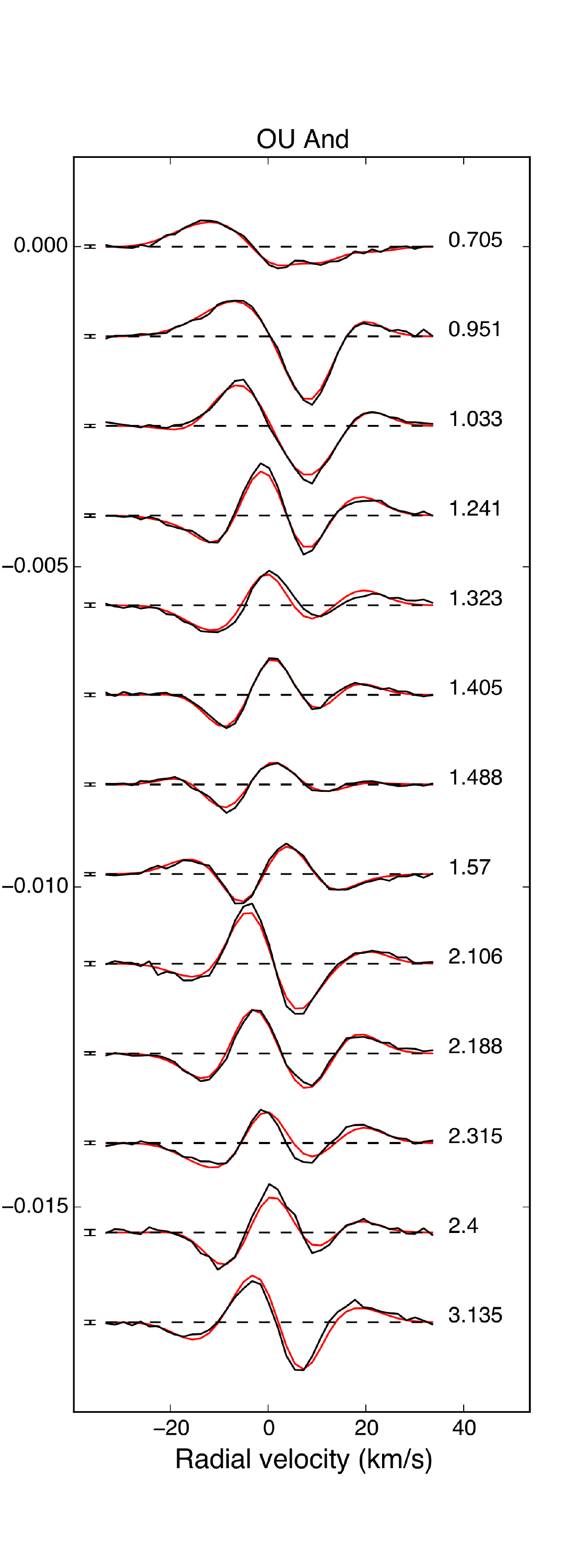}
\includegraphics[width=0.85\columnwidth]{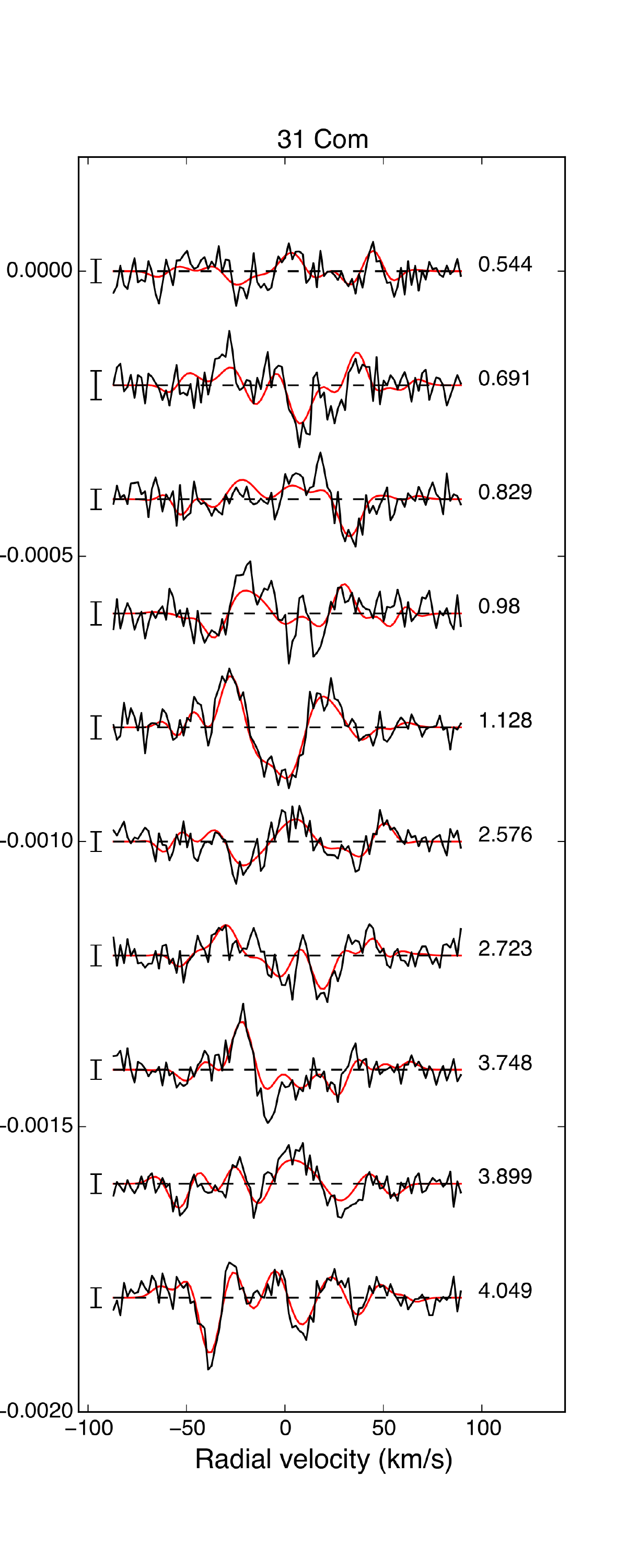}
\caption{Observed Stokes~V LSD profiles (black line) and profiles produced by our best ZDI model (red lines) for OU~And (left) and 31~Comae (right). Successive profiles are vertically shifted (from top to bottom) for display clarity. The rotational phase of each observation (computed from the ephemeris given in the text) is listed to the right, with the integer part denoting the rotational cycle since the start of the observing run. Error bars are illustrated on the left of every profile and dashed lines illustrate the zero level.}
\label{fig:stokesv}
\end{figure*}

\begin{figure*}
\centering
\includegraphics[width=1.0\columnwidth]{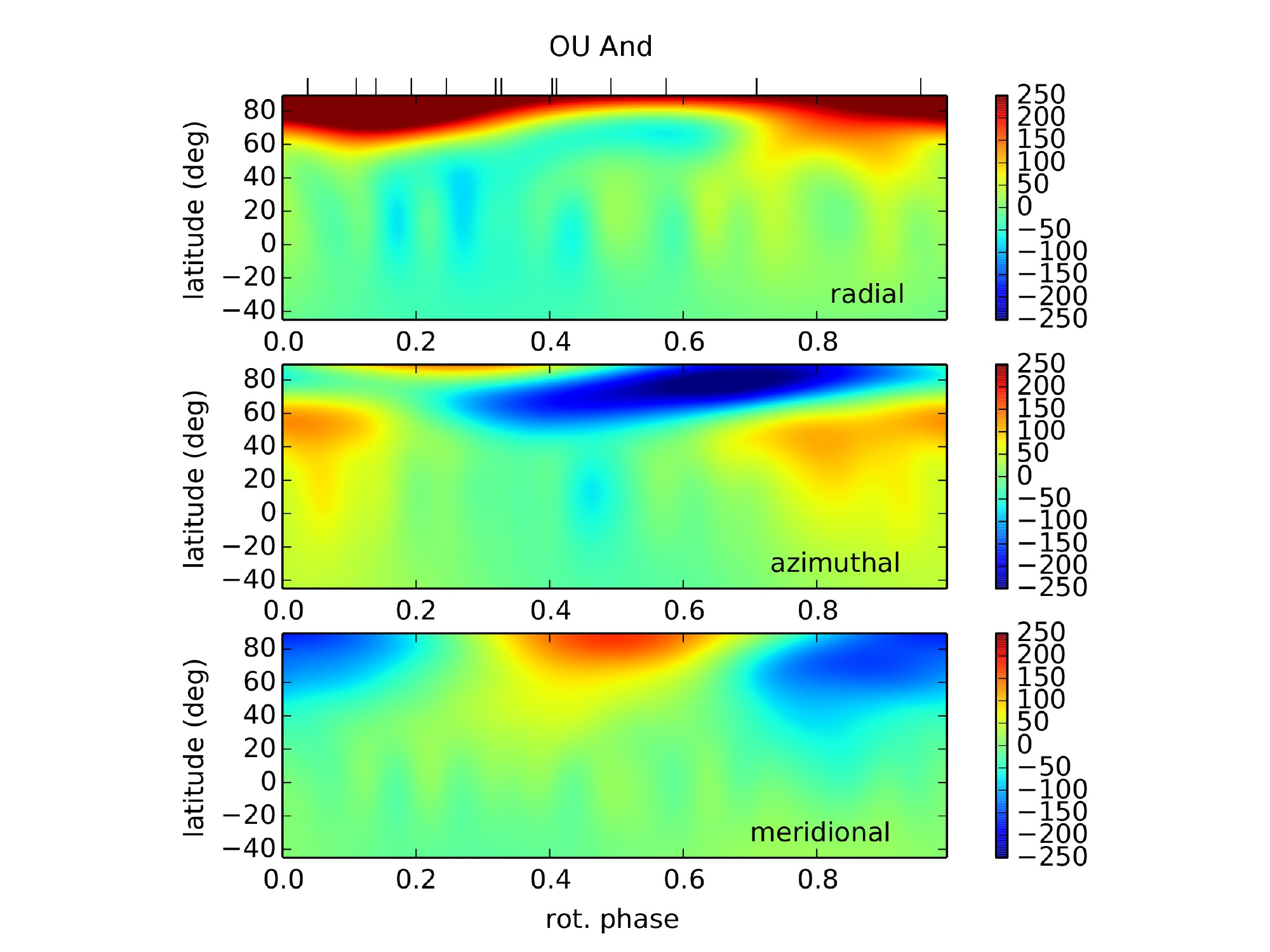}
\includegraphics[width=1.0\columnwidth]{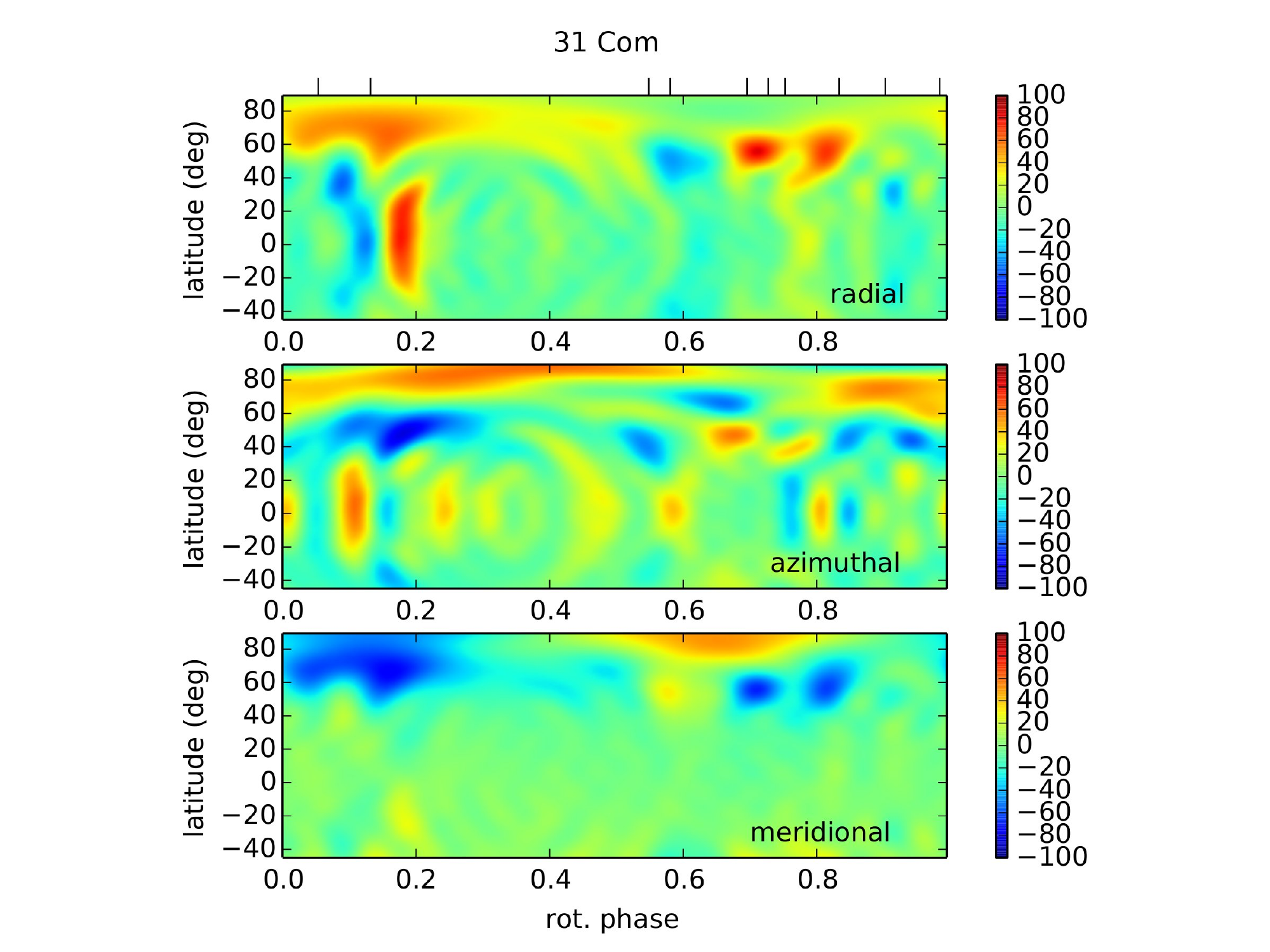}
\caption{Photospheric magnetic maps of OU~And (left) and 31~Com~(right) in 2013. For each star, the vectorial magnetic field is split in its three components in spherical projection and $\ell_{\rm max}=20$. The color scale illustrates the field strength in Gauss, and the rotational phases of our data sets show up as vertical ticks on top of each map.}
\label{fig:maps}
\end{figure*}

\begin{figure*}
\centering
\includegraphics[width=0.85\columnwidth]{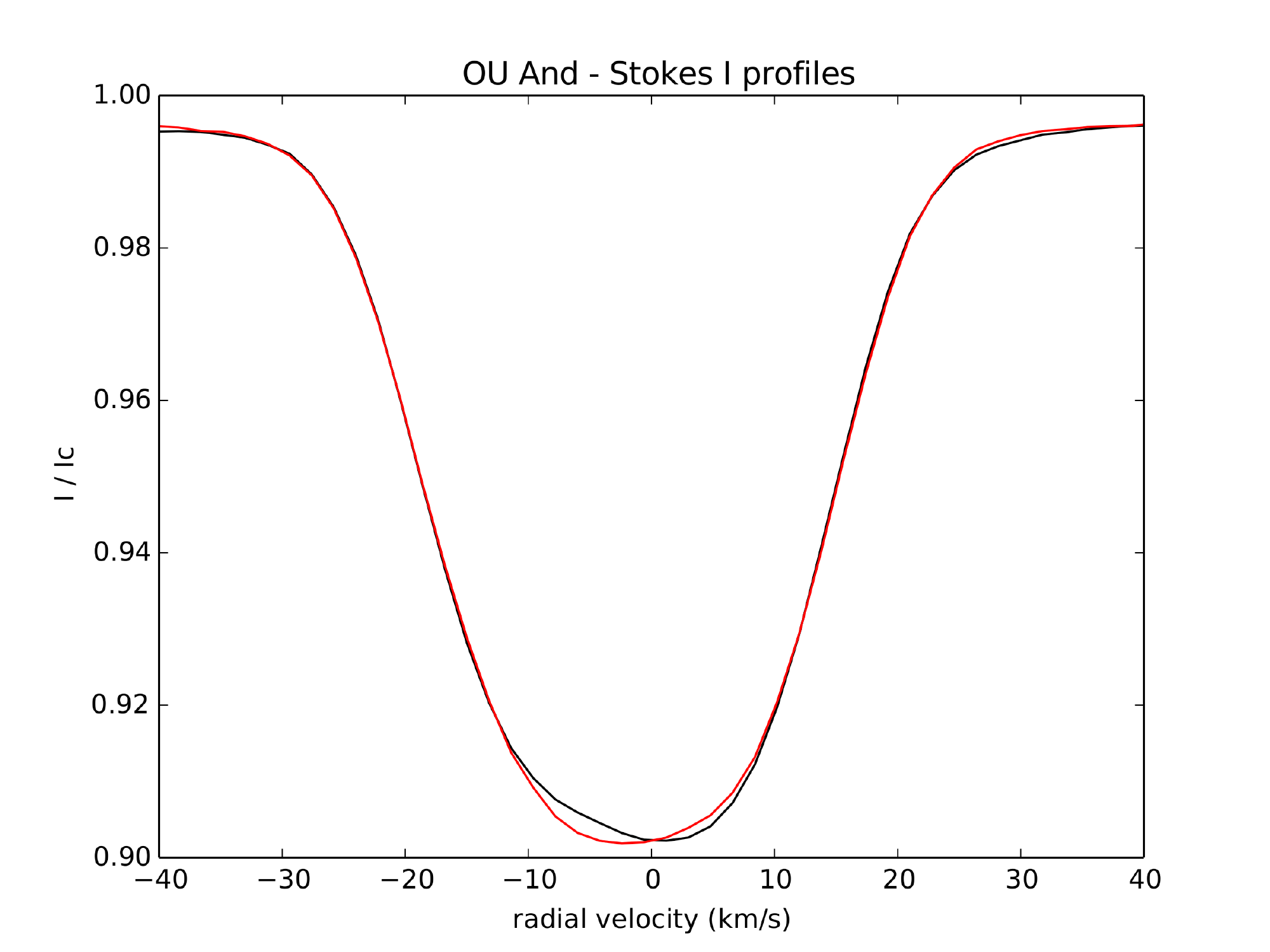}
\includegraphics[width=0.85\columnwidth]{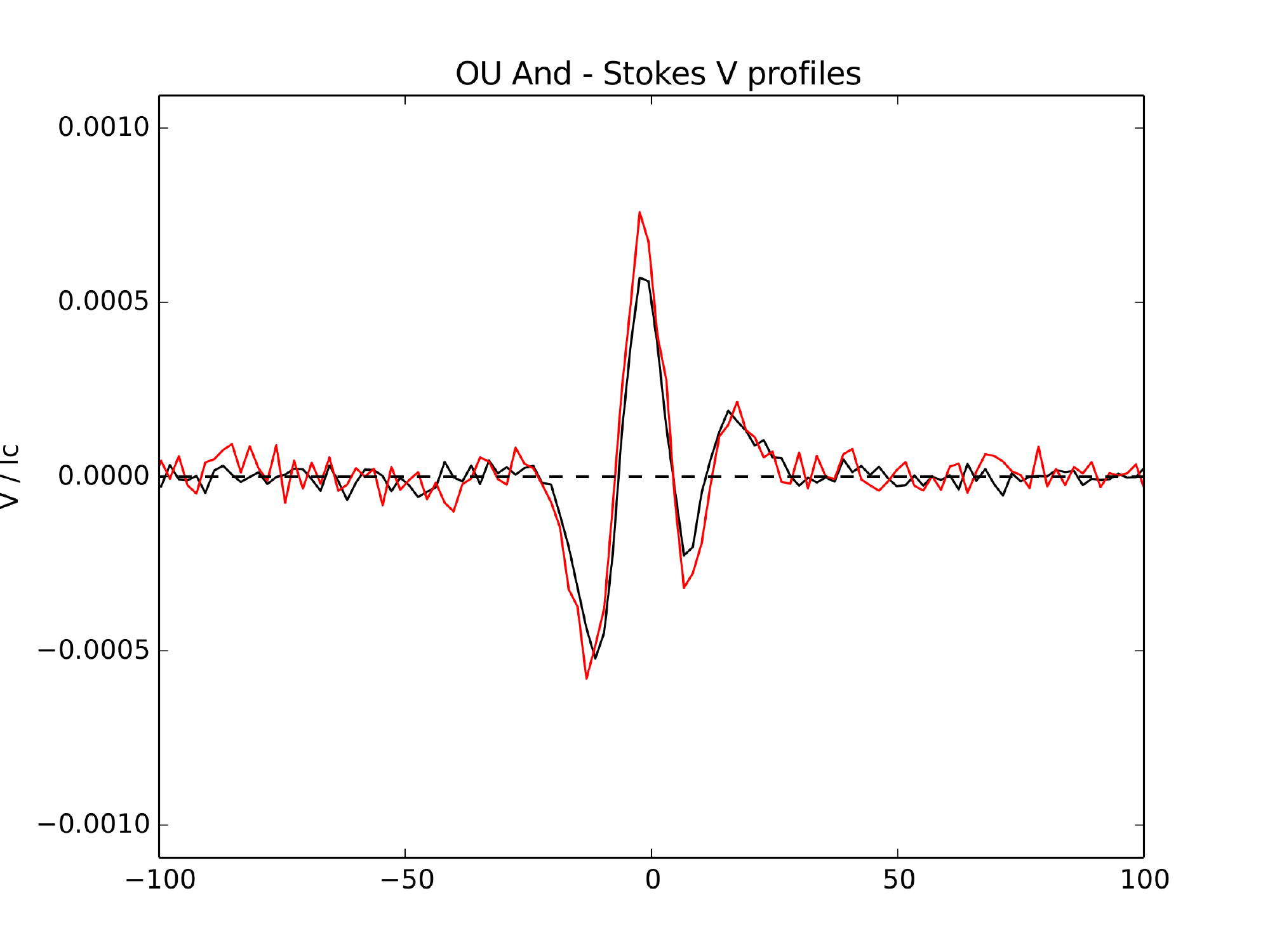}
\caption{Evolution of Stokes~I and Stokes~V (left and right, respectively) LSD profiles of OU~And observed close to phase 0.4. We represent in black the observation of 2013 Sept 19 (phase 0.41) and in red the observation of 2013 Oct 13 (phase 0.40).}
\label{fig:2profiles}
\end{figure*}

The temporal series of observations gathered for both stars offer the opportunity to obtain polarized measurements at various rotational phases. Using this multi-angle view of the stellar photosphere, tomographic methods can be used to model the surface magnetic geometry of the targets.

\subsection{Zeeman-Doppler imaging}
\label{subsection:ZDI}
The tomographic algorithm employed here is based on the Zeeman-Doppler
imaging (hereafter ZDI) method proposed by \cite{semel89}. More
specifically, we use the maximum-entropy ZDI code described by
\cite{donati_2006}. The ZDI principle makes use of the fact that the
Doppler broadening of the polarized profiles results in a 1D resolution of
the stellar surface (orthogonally to the rotation axis), as magnetic spots
located at different places on the visible stellar hemisphere possess
different radial velocities and, therefore, produce polarized spectral
signatures showing up with a different wavelength shift with respect to the
line center. The succession of observations gathered at different
rotational phases is then used to recover the full surface distribution of
the local magnetic vector, by iterative adjustment of the time-series of
LSD profiles.

The synthetic LSD profiles produced by ZDI rely on a simplified atmospheric
model, based on a spherical artificial star split into a grid of pixels of
roughly identical areas. Each pixel is attributed a brightness that we
assume to be constant over stellar surface (but affected by limb
darkening), and the elemental areas host a magnetic field with a given
strength and orientation. Each pixel produces Stokes~$I$ and $V$ spectral
lines, computed under the weak-field approximation (which remains valid up
to a few kG, \citet{kochukhov_2010}). In the weak-field regime, the local
Stokes I spectral line is unaffected by the magnetic field and we crudely
assume here that the line profile possesses a Gaussian shape. The Stokes V
line profile is proportional to the derivative of the Stokes~$I$ profile and
to the line-of-sight projection of the local magnetic vector magnetic
field. It is also proportional to the effective Land\'e factor of the LSD
pseudo-line profile and to the square of its effective wavelength.

The present implementation of this method includes the projection of the
magnetic field onto a spherical harmonics frame. Compared to earlier
studies based on a stellar surface divided in a grid of independent pixels,
the spherical harmonics frame has the main advantage to offer an easy
splitting of the field into its poloidal and toroidal components \citep{donati_2006}, 
providing us with a practical separation of these two
important surface outputs of stellar dynamos. The other advantage of
employing spherical harmonics is the better behavior of the code for
low-degree magnetic geometries \citep{donati_2001}, while older versions of the
code were reported to perform poor reconstructions of simple magnetic
configurations \citep[e.g. inclined dipoles, ][]{donati_1997}.

\subsection{Magnetic mapping of OU~And}
 \label{subsection:MMOUAnd}
The 13 rotational phases collected for OU~And in 2013 provide us with a good basis
to calculate a ZDI model, in spite of a paucity of observations
between phase 0.7 and 0.95. Another asset of the data set is its low noise
level (relative to the amplitude of the polarized signal). Using $v\sin(i)= 21.5$~km.s$^{-1}$, 
$i = 50^\circ$, $P_{rot}$ = 24.2~d, and using all spherical
harmonics modes with $\ell \le \ell_{\rm max}=20$, the ZDI procedure leads
to a reduced $\chi^2$ (hereafter $\chi^2_r$) of 3, showing that the
accuracy of data fitting remains significantly worse than the noise level. The
comparison of observed versus synthetic Stokes V profiles illustrated in
Fig.~\ref{fig:stokesv} clearly shows that some rotational phases give
reasonably well reproduced polarized profiles (e.g. phase 0.951), while
other modeled Stokes V signatures display a significant departure from their
observational counterpart (e.g. the last observation, at phase 3.135).

A first possible reason for our difficulty in fitting the data
may be the use of a wrong rotation period to compute the rotational
phases of our time-series. Here, we adopted  the photometric
period of 24.2 d reported by \cite{strass_99}, and a
naked-eye estimate is enough to see that this value is consistent
with the variability of the Stokes V LSD profiles, considering the
similar shape of polarized profiles obtained at nearby phases, one
rotation cycle apart (e.g. phases 1.32 and 2.32). To refine the
period value, we ran a period search by performing a number of
ZDI models with all input parameters left unchanged, except the
rotation period that we vary from 15 to 30 d. Using the maximum
entropy principle, the most likely model among this grid
is the one that minimizes the average field strength of the map
at a fixed $\chi^2$, where we use the average surface field as a proxy of the information content of the map. Equivalently, the best model is
also the one that minimizes the $\chi^2$ at a fixed information content.
By doing so, we strictly reproduce the approach successfully
adopted by \cite{auriere_2011} or \cite{tsvetkova_2013} for other active giants. For OU And, the optimal period found
with this strategy using data from 2013 is very close to the one proposed by Strassmeier, with a ZDI value of 24.13~d, giving further confirmation of the robustness of the period choice. If we use this slightly modified
period to reconstruct a new map, we actually find a marginal
improvement of the model compared to our initial one, and no
noticeable change in the field distribution obtained as an output. We then repeated this test after combining our data from 2008 and 2013 in a single time-series,  to estimate the benefit of a larger timespan. The main outcome of this second test is very degraded $\chi^2$ value (as high as 17) implying that, over such longer period of time, pure rotational modulation is not the only contributor to the variability of Stokes V. This time, we are not able to identify a clear minimum, presumably because of the poor quality of the ZDI models. We therefore use the 24.2~d period in the rest of our work, since it was determined using photometric data with a larger span of time than our observations.

We also reconstructed a number of ZDI maps based on different values of the inclination angle, which highlighted a best fit for $i = 50^\circ$. This inclination value is somewhat at odds with the independent estimate that can be obtained by combining the values of the stellar radius, $v\sin i$ and rotation period, pointing towards an inclination angle close to $90^\circ$. In previous ZDI studies of active giants, this approach was able to provide us with a value for the inclination that was consistent with other stellar parameters \citep{auriere_2011,tsvetkova_2013}. The origin of the discrepancy obtained for OU And is unclear. We note however that a ZDI reconstruction with i greater than $75^\circ$ results is a very high $\chi^2$ value, suggesting that an equator-on configuration is very unlikely. We therefore adopt $i = 50^\circ$ in all ZDI models discussed hereafter.

Another possible origin of the mismatch between observations and synthetic
profiles is the duration of the observing run, which contains data
collected over nearly 2 months. This time-span may be long enough to allow
for a temporal evolution of the field geometry and therefore break the
basic assumption of ZDI that rotation alone is responsible for Stokes V
variability. Temporal changes are actually observed in both Stokes I and V
line profiles observed at nearby rotational phases, but at least one
rotation cycle apart from each other. A good illustration comes from data
obtained at phases 1.405 and 2.400 (Fig.~\ref{fig:2profiles}), for which
Stokes V profiles display a roughly similar shape, but with a slightly
larger amplitude in the most recent observation. The situation is more
critical for Stokes I, for which the positive bump seen in the blue side of
the LSD profile core in September is no more observed one month later and is
replaced by a similar bump in the red side of the line core. This
evolution, which is apparently taking place faster in brightness inhomogeneities
than in polarized signatures, resulted in convergence problems in an
attempt to reconstruct a Doppler map (not shown here) from our set of
Stokes I LSD profiles.

Specific types of surface changes can be taken into account in the
inversion procedure, as long as they are systematic. This is the case, for
instance, for latitudinal differential rotation that can be taken care of
in the inversion procedure using a simple latitudinal dependence of the
surface shear \citep{petit_2002}. In spite of a reasonably dense
rotational sampling, our ZDI model, which incorporates differential rotation, does
not highlight a unique set of differential rotation parameters to be able to further optimize  the magnetic model. 
This may be due to the fact that the very subtle changes in Stokes V generated by differential rotation are easily dominated
by other types of variability of the surface magnetic geometry, in particular the emergence and decay of magnetic patches. Table~\ref{tab:dynamo} (middle lines) gives the parameters for the magnetic models obtained for OU~And in 2013 with the same parameters as used for Fig.~\ref{fig:maps}, with $\ell_{\rm max}=10$ and $\ell_{\rm max}=20$. The reconstruction of ZDI models with different $\ell_{\rm max}$ values is meant to help in the comparison of magnetic maps of stars affected by a very different $v\sin(i)$, which leads to a very different spatial resolution of the stellar surface through tomographic inversion (e.g. Morin et al. 2010). While $\ell_{\rm max}=20$ is necessary for capturing the complexity of the Stokes~V signatures of 31 Com in our model, $\ell_{\rm max}=10$ is actually enough at the lower $v\sin(i)$ of OU And. In this context, the apparently more complex field geometry of 31 Com is, at least partially, due to the better surface resolution through ZDI. Reconstructing a ZDI model of 31 Com 
limited to $\ell_{\rm max}=10$ is therefore a way to impose a low-pass spatial filtering that will compensate for this effect (at the cost of a higher $\chi^2_r$). Figure~\ref{fig:maps} (left) illustrates the magnetic map obtained for $\ell_{\rm max}=20$. The magnetic field topology is mainly poloidal (64 \% of the total magnetic energy) and the poloidal component is mainly dipolar (58 \% of the magnetic energy in the poloidal component). Figure~\ref{fig:2profiles} clearly shows that the map is dominated by large-scale features, although the rather high $v\sin(i)$ would enable smaller structures to be mapped. In particular, 
the positive pole of the magnetic dipole dominates the radial component, when a toroidal component of negative polarity appears clearly in the azimuthal component.

A second spectropolarimetric data set was previously obtained in 2008. This
time series covers only 15 nights, so that about one third of the stellar rotation
remains uncovered at this epoch. Only six rotation phases are available at that time
for a total of nine spectra, four observations being gathered on September 16. Polarized
signatures are detected in all LSD profiles, and this material was used to
reconstruct a second magnetic map of OU~And (based on the rotation period of 24.2~d), which was derived from our observations in 2013, since the phase coverage is not dense enough to permit an independent period search from the 2008 data alone.

\begin{figure}
\centering
\includegraphics[width=0.70\columnwidth]{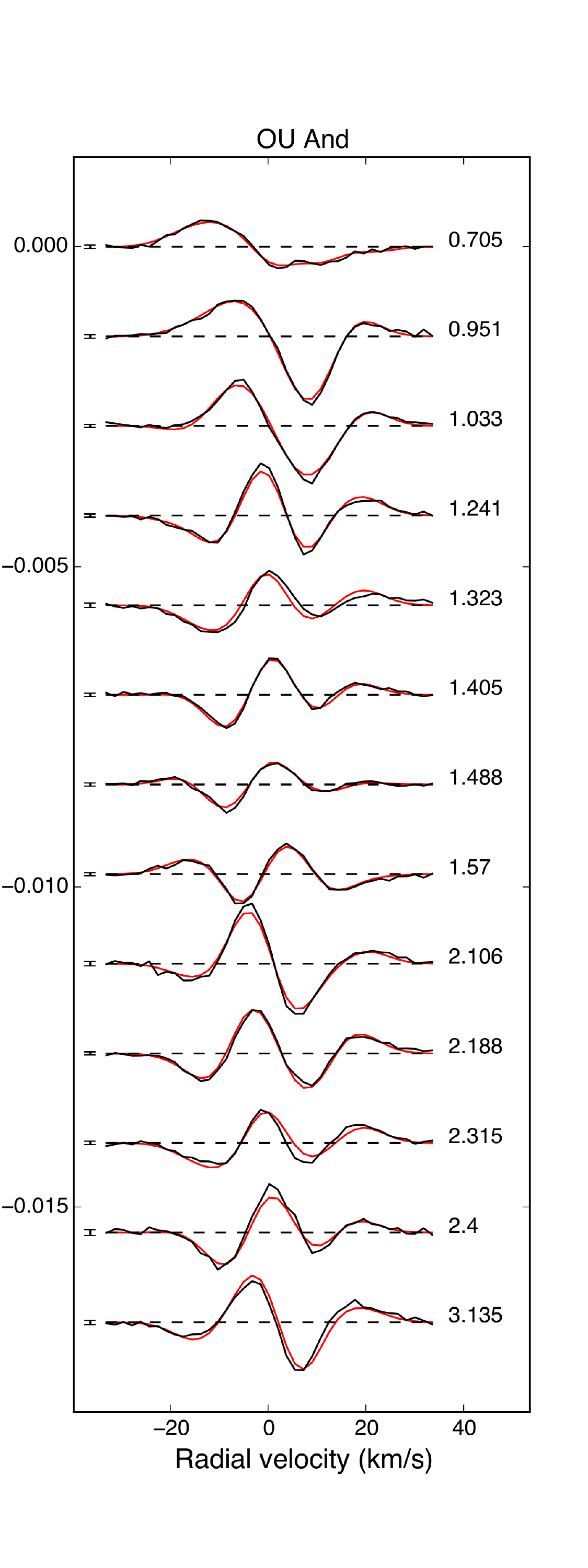}
\centering
\includegraphics[width=0.95\columnwidth]{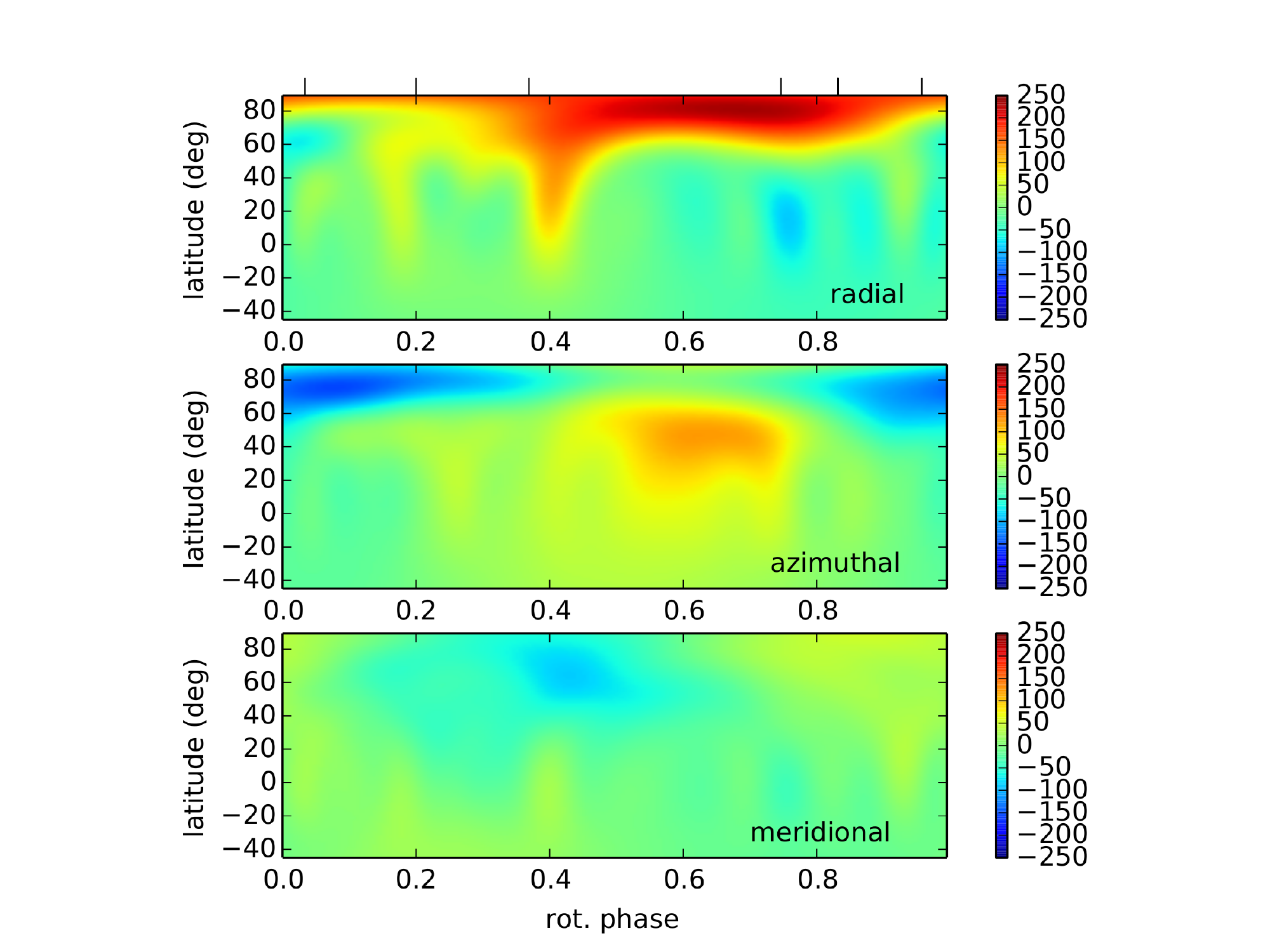}

\caption{Stokes V LSD profiles (upper panel) and the surface magnetic map (bottom panel) for observations of OU~And gathered in 2008 with $\ell = 20$. The plots follow the same definitions as in Figs.~ \ref{fig:stokesv} and~\ref{fig:maps}.}
\label{fig:ouand08_a}
\end{figure}

The magnetic model extracted from 2008 data is presented in the lower panel of Fig.~\ref{fig:ouand08_a}. The values of
input ZDI parameters are identical for 2008 and 2013. The $\chi^2_r$ value for this
new map reconstruction is equal to 0.9. Magnetic quantities extracted from the new
map (Tab.~\ref{tab:dynamo}) show that the average field strength over stellar surface is slightly weaker than in 2013, the same being observed for the maximal field strength. The
sparse phase coverage available in 2008 may be (at least partially) responsible for
this limited difference. Other general characteristics of the field geometry are
remarkably similar at both epochs, with a simple field structure dominated by a
dipole. 

\begin{figure}[htb]
%\centering
\includegraphics[width=0.8\columnwidth]{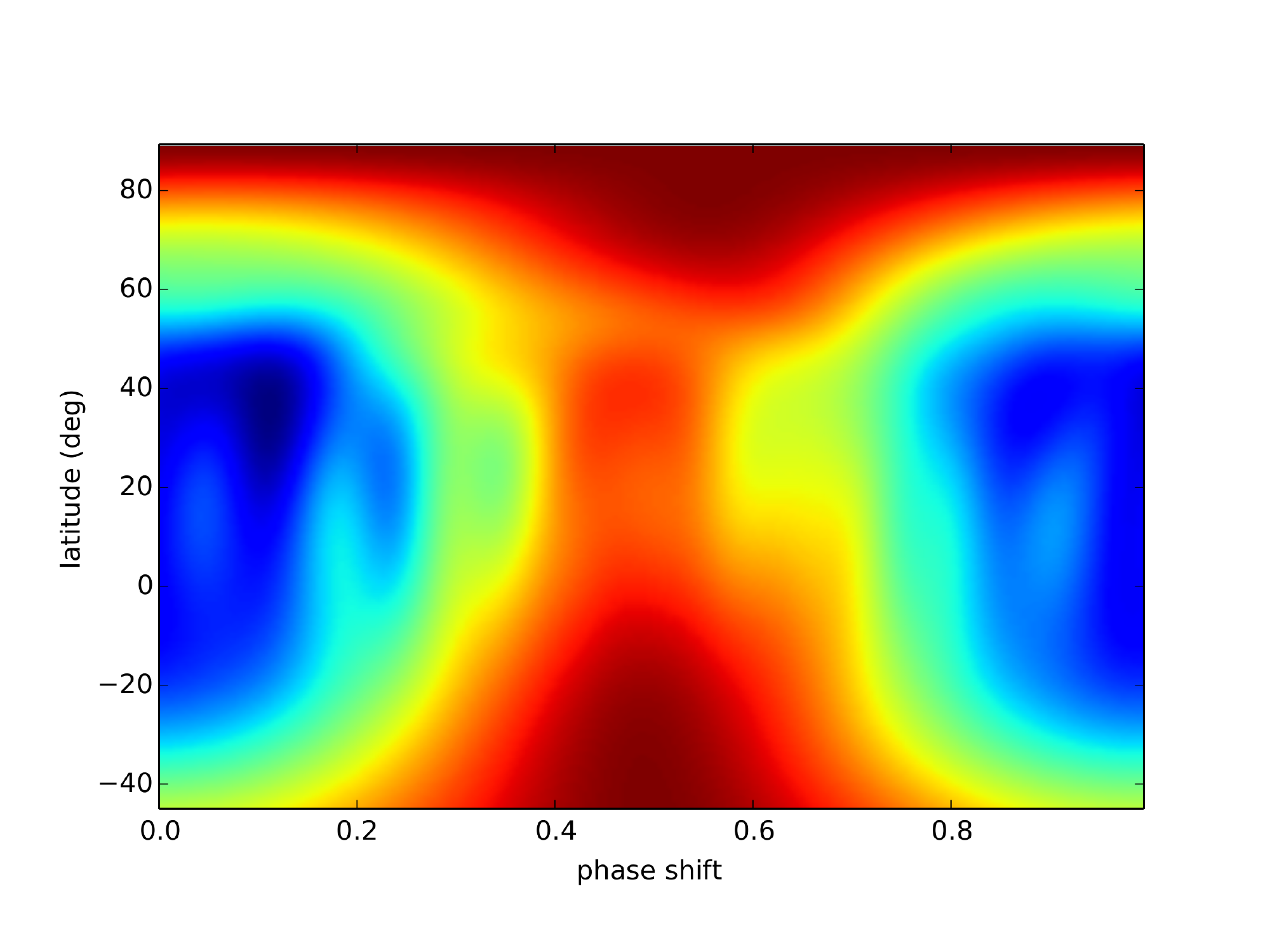}
\includegraphics[width=0.8\columnwidth]{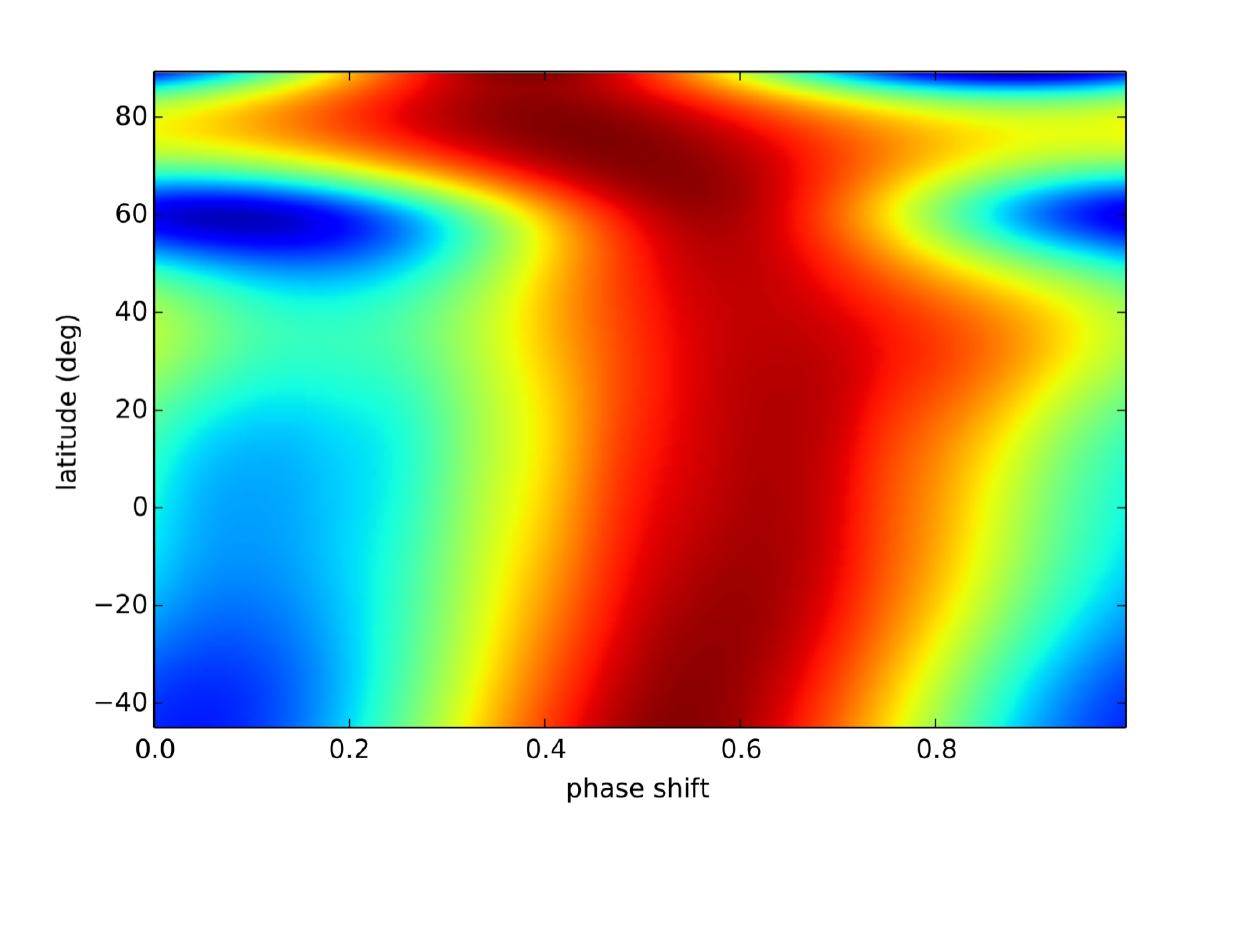}
\includegraphics[width=0.815\columnwidth]{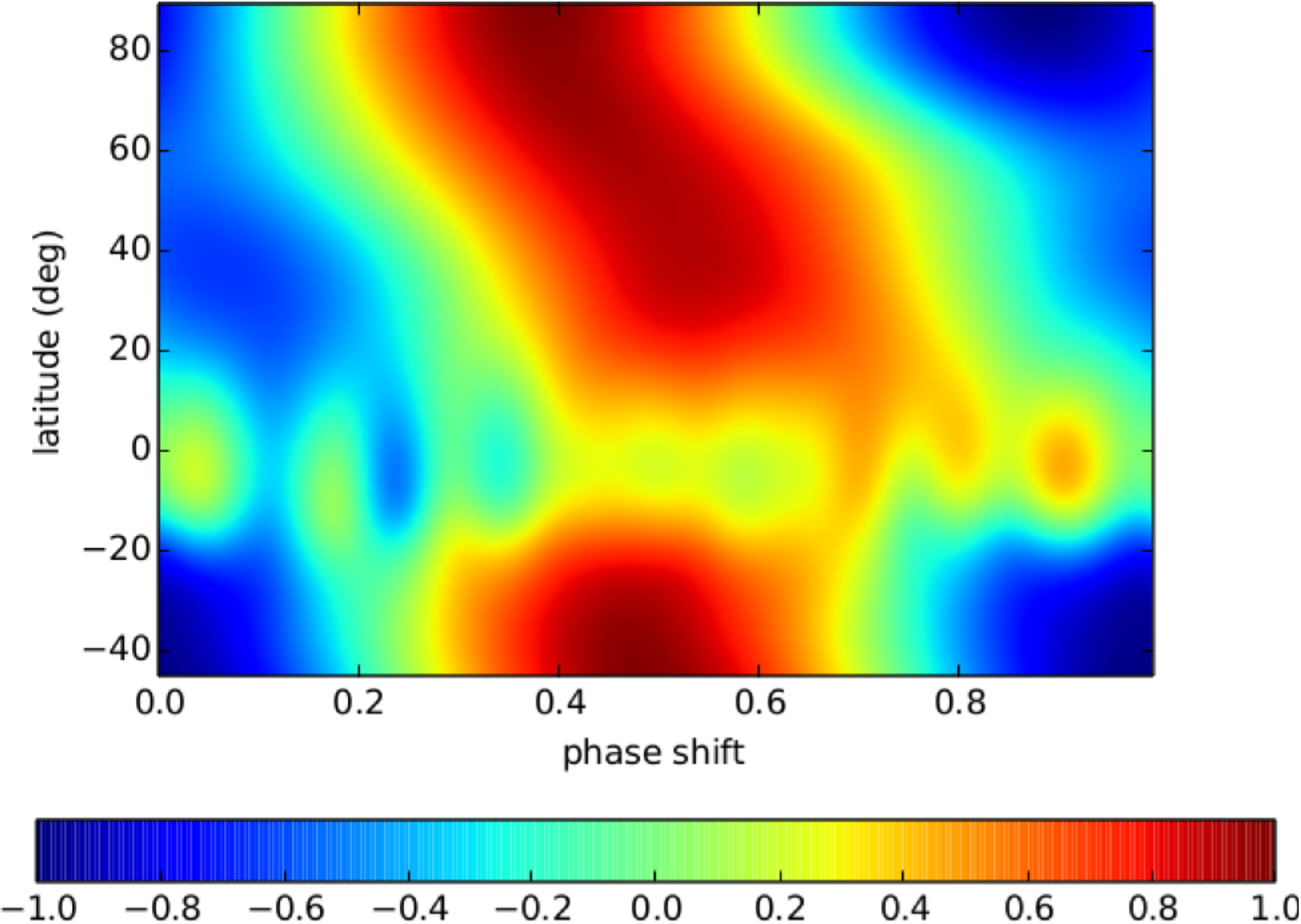}
\caption{Cross-correlation (2008 versus 2013) of magnetic maps of OU~And derived for all three components of the magnetic vector with, from top to bottom, the radial, azimuthal, and meridional field projection.}
\label{fig:crosscor}
\end{figure}

To perform a more systematic comparison of the magnetic maps obtained five years apart,
we cross-correlated every latitudinal band of the two maps following, for example, \citet{petit_2010}, for each field component separately, which resulted in the three
cross-correlation maps illustrated in Fig.~\ref{fig:crosscor}. Using this approach, we highlight a strong correlation of both maps at all latitudes, provided that a $\approx 0.5$ phase shift is applied. This correlation is easily visible by comparing the 2008 and
2013 maps with the naked eye, noticing that the maximal radial field value is observed
at around phase 0.65 in 2008, and 0.15 in 2013. The lack of observations between
2008 and 2013 is obviously too large to decide whether this overall similarity is
due to a global stability of the field geometry over this timespan, in which case
the 0.5 phase shift between both maps would be compatible with our relatively large
uncertainty on the stellar rotation period.

\subsection{Magnetic mapping of 31~Com}
 \label{subsection:MM31Com}
According to the 6.8~d rotation period proposed by \cite{strass_2010}, 
the time series available for 31~Com is affected by a large
phase gap, which spans about 40\% of the rotation cycle, between phases
0.13 and 0.54. Given the relatively high inclination angle of the
stellar spin axis (of the order of 80$^\circ$ when the rotation
period, $v.\sin(i)$ and radius are considered together, in the solid-body rotation hypothesis), this sparse phase coverage implies that a significant fraction of the
theoretically accessible stellar surface remains completely
unobserved for phases inside the gap and stellar latitudes below
80$^\circ$. In spite of this well identified limitation, we decide to
use our ten observations to reconstruct the large-scale magnetic
geometry of its photosphere through ZDI.

For this we use  $v\sin(i)=67$~km.s$^{-1}$ and $i=80^\circ$. Because of
the relatively high $v.\sin(i)$ value, we use a model with more spherical harmonics $\ell_{\rm max}=20$, to take full advantage of the higher spatial resolution offered by a larger projected rotational velocity. We also compute a ZDI model restricted
to $\ell_{\rm max}=10$, to ease the comparison with our ZDI model of
OU~And. A set of ZDI models was computed for various values of the
rotation period (between 2~d and 10~d), but the outcome does not
reveal a preferred rotation period, presumably because of  the poor phase
coverage. A systematic search for latitudinal differential rotation
was similarly unsuccessful. We therefore keep the period of \cite{strass_2010} 
to compute the rotation phases, and assume solid-body rotation in the ZDI procedure.

The magnetic map resulting from the tomographic procedure with
$\ell_{\rm max}=20$ is illustrated in Fig.~\ref{fig:maps} (right), while the fit to the
observed Stokes~V profiles is shown in Fig.~\ref{fig:stokesv}. Magnetic quantities
extracted from the maps are listed in Table~\ref{tab:dynamo}. We achieve here
$\chi^2_r=1.2$, showing that our model is able to reproduce the set of
observational data close to the level of photon noise. The surface map
displays a complex field geometry, with a majority (69\%) of the
poloidal magnetic energy stored in spherical harmonics modes with
$\ell > 3$. The surface magnetic energy is also equally distributed
between the poloidal and toroidal field components. Owing to the high
inclination angle, a majority of both stellar hemispheres can be
observed. The apparent symmetry of the reconstructed magnetic geometry
about the equator  actually illustrates the difficulty of the
tomographic code to distinguish between the hemispheres in this
specific geometrical configuration. As expected with our sparse phase
coverage, we note an absence of reconstructed magnetic regions between
phase 0.2 and 0.5, owing to the lack of observational constraints at these phases. 
A likely consequence of this situation is that the average field strength reported in Table~\ref{tab:dynamo} may be
underestimated by about 30\%, because a large phase gap is hiding a significant fraction of the stellar surface and the maximum entropy reconstruction will force a null magnetic field in this unseen region.

Contrary to OU~And for which maps with $\ell_{\rm max}=10$ and
$\ell_{\rm max}=20$ can barely be distinguished (as illustrated by the
numbers reported in Table~\ref{tab:dynamo}), limiting the spherical harmonics
expansion to $\ell_{\rm max}=10$ in our model of 31~Comae induces
measurable modifications in the magnetic geometry. In particular, we
observe that the surface smoothing imposed through $\ell_{\rm max}=10$
reduces the maximal field strength by about 7\%, and increases by
about 20\% the amount of the magnetic energy reconstructed in modes
with $\ell < 4$.
\begin{table*}
\caption[]{Magnetic parameters.}
\centering
\begin{tabular}{lcccccccc}
\hline
Star                 & $\ell_{\rm max}$ & $B_{\rm mean}$ & $B_{\rm max}$   & pol. en. & dipole   & quad.     & oct.         & axi.          \\
                     &                  & (G)            & (G)             & (\% tot) & (\% pol) & (\% pol)  & (\% pol)     &  (\% tot) \\
\hline
OU~And (2008)        & 20               & 57             & 235             & 82       & 56       & 14        & 10           & 40     \\
OU~And (2008)        & 10               & 57             & 234             & 82       & 57       & 14        & 10           & 40     \\
\hline
OU~And (2013)        & 20               & 68             & 392             & 64       & 58       & 10        & 8            & 34     \\
OU~And (2013)        & 10               & 68             & 383             & 64       & 58       & 10        & 9            & 34     \\
\hline
31~Com               & 20               & 32             & 155             & 56       & 9        & 15        & 7            & 25     \\
31~Com               & 10               & 30             & 143             & 58       & 11       & 18        & 8            & 28     \\
\hline
\end{tabular}
\\
\noindent Notes: Magnetic quantities derived from the set of magnetic maps. For every map, we list the maximal $\ell$ value of the spherical harmonics projection ($\ell_{\rm max}$), the mean unsigned magnetic field ($B_{\rm mean}$), and the maximal field strength over stellar surface ($B_{\rm max}$). We then give the fraction of the large-scale magnetic energy reconstructed in the poloidal field component and the fraction of the poloidal magnetic energy in the dipolar ($\ell = 1$), quadrupolar ($\ell = 2$) and octopolar ($\ell = 3$) components, as well as the fraction of the magnetic energy in the axisymmetric field component ($m = 0$).
\label{tab:dynamo}
\end{table*}

\section{Evolutionary status}  %5
\label{section:evolution}
  \subsection{Stellar parameters}
  \label{subsection:stellar_parameters}
  
For OU~And, $T_{\rm eff}$ = 5850~K is inferred by \cite{wright_2003} from its spectral type. \cite{gondoin_2003, gondoin_2005b} uses a cooler temperature (5110~K) but do not indicate how it was obtained. 
We then made our own determination of atmospheric parameters, as presented in Table~\ref{tab:stars}, and used derived values in this work as well as in \cite{big_paper}. 
We determined parameters of OU~And, such as effective temperature ($T_{\rm eff}$), surface gravity ($\log g$), microturbulence ($\xi$), and metallicity using the local thermodynamic equilibrium (LTE) model atmospheres of \cite{Kurucz_18CD} and the current version of the spectral analysis code {\sc moog} \citep{moog}. We measured equivalent widths (EW) of 76 \ion{Fe}{i} and 6 \ion{Fe}{ii} lines in the spectrum obtained on September 25, 2008. The $\log gf$ values were taken from \cite{lambert_96}. 
The solution of the excitation equilibrium used to derive the effective temperature ($T_{\rm eff}$) 
was defined by a zero slope of the trend between the iron abundance derived from \ion{Fe}{i} lines 
and the excitation potential of the measured lines. 
The microturbulent velocity ($\xi$) was found by constraining the abundance, determined from individual 
\ion{Fe}{i} lines, to show no dependence on $EW_\lambda/\lambda$. 
The value of $\log g$ was determined by means of the ionization balance assuming LTE. 
We then found $T_{\rm eff}=5360$~K and $\log g=2.80$. Typical uncertainty in $T_{\rm eff}$ and 
$\log g$ are $\pm$120~K and $\pm$0.2, respectively.

Using the above values for $T_{\rm eff}$, $\log g$, and luminosity, the radius of the star 
is estimated to be $R =9.46\,R_{\odot}$ assuming the bolometrical correction BC=$-$0.16 (\cite{alonso_99} and the reddening $A_V=0.09$. The photometric period of OU~And varies in different observational epochs. For magnetic field mapping, we adopted the value of 24.2~d that is close to the best fit of the ZDI models for 2008 and 2013 data, which was treated separately.

31~Comae is a member of the open cluster Mellote~111, also known as Coma Berenices Cluster 
\citep{casewell_2006} and we may assume the metalicity and reddening follow those of the cluster. 
Mellote~111 has solar metalicity [Fe/H]=0.00$\pm$0.08 \citep{heiter_2014} and its age is about 
$560\pm 90$~Myr \citep{silaj_2014}. Interstellar reddening in the direction of the Mellote~111 
is $E(B-V)=0.00$ \citep{robichon_99}. Because of the very fast rotation of 31~Com, the lines in its 
spectrum are wide and blended and it is impossible to determine atmospheric parameters of this
star in the same way as we did for OU~And, measuring EWs of individual \ion{Fe}{i} and \ion{Fe}{ii} 
lines. For this reason we calculated a grid of synthetic spectra in a wide range of effective 
temperatures ($T_{\rm eff}$) in the spectral region 6080 - 6180~\AA, which contains lines of neutral 
elements with different excitation potentials and lines of \ion{Fe}{ii,} which are sensitive to 
surface gravity.
Synthetic spectra calculated with $T_{\rm eff}$ from 5550~K to 5700~K fit well 
the observed spectrum and it is difficult to obtain $T_{\rm eff}$ with higher precision.
Additionally, photospheric spots and active regions contribute to the strengthening or weakening 
of the lines of different elements. To derive the $T_{\rm eff}$ value with higher precision, we used also photometric calibrations (see Table \ref{tab:T_31Com}).

\begin{table*}
\caption[]{Photometric calibrations of the effective temperature of 31~Com.}
\centering
\begin{tabular}{lccc}
\hline
\hline
Colour  &       & $T_{\rm eff}$,~K & References \\
\hline
$(B-V)$  & 0.67  & 5521 & (1), (2) \\
$(J-Ks)$ & 0.369 & 5682 & (1), (3) \\
$(B-V)$  & 0.67  & 5562 & (4), (2) \\
$(b-y)$  & 0.435 & 5536 & (4), (5) \\
$(J-H)$  & 0.30  & 5644 & (4), (2)  \\
$(J-K)$  & 0.37  & 5601 & (4), (2)  \\ 
$(R-I)$  & 0.35  & 5558 & (4), (2)  \\
\hline
\end{tabular}
\tablebib{(1)~\cite{gonzales_2009}; 
          (2)~\cite{ducati_2002};
          (3)~\cite{2MASS};
          (4)~\cite{alonso_99};
          (5)~\cite{Hauck_98};
           }
\label{tab:T_31Com}
\end{table*}

We also found  the following estimations of the effective temperature of 31~Com in the literature:
5689~K \citep{katz_2011}, 5747~K \citep{blackwell_98}, 5761~K \citep{houdashelt_2000},
5623~K \citep{massarotti_2008}, and 5660~K \citep{strass_2010}. 
Further in the paper we will use this latter value of the effective temperature
($T_{\rm eff}=5660$~K), which we also used  in Auriere et al. (2015).
For radial velocity, rotation period, and $v\sin(i),$ we  adopted the values derived by \cite{strass_2010}.
We calculated the luminosity $L=73.4~L_\odot$ and stellar radius of $R=8.5~R_\odot$ using BC=$-$0.114 \citep{alonso_99} and $E(B-V)=0.00$. Stellar mass $M=2.75 M_{\odot}$ and surface gravity $\log g =2.97$ are based on the evolutionary tracks 
discussed in the next section. Higher value of $\log g =3.51$ for 31~Comae, as suggested by \cite{strass_2010}, does not match our evolutionary models.

  \subsection{Evolutionary status, mass, and rotation}
  \label{subsection:evolution}
  
\begin{figure}[htb]
\centering
\includegraphics[width=1.0\columnwidth]{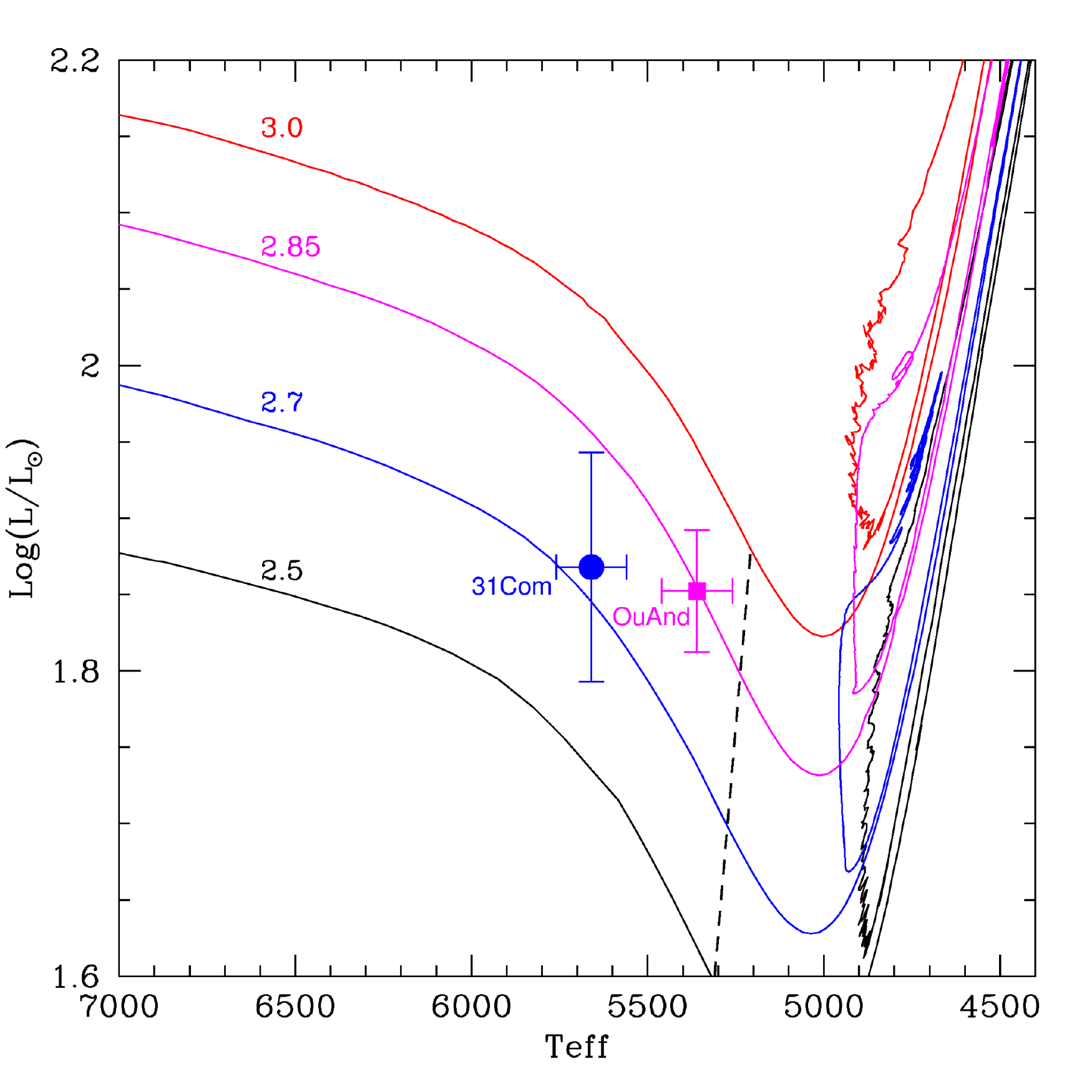}
\caption{Position of 31~Comae and OU~And (circle and square, respectively) on the Hertzsprung–Russell diagram (HRD)
diagram. Solar metallicity tracks of evolution models with different 
masses (indicated on the plot, in solar mass) are shown. The dashed 
black line delimits the beginning of the first dredge-up phase and 
corresponds with the evolution points when the mass of the convective 
envelope of the models encompasses 2.5\% of the total stellar mass. 
The loops on the evolutionary tracks in the right part of the HRD 
(effective temperature below $\sim$ 5000~K) correspond to the central helium-burning phase.}
\label{fig:evolution}
\end{figure}

\begin{figure}[htb]
\centering
\includegraphics[width=0.9\columnwidth]{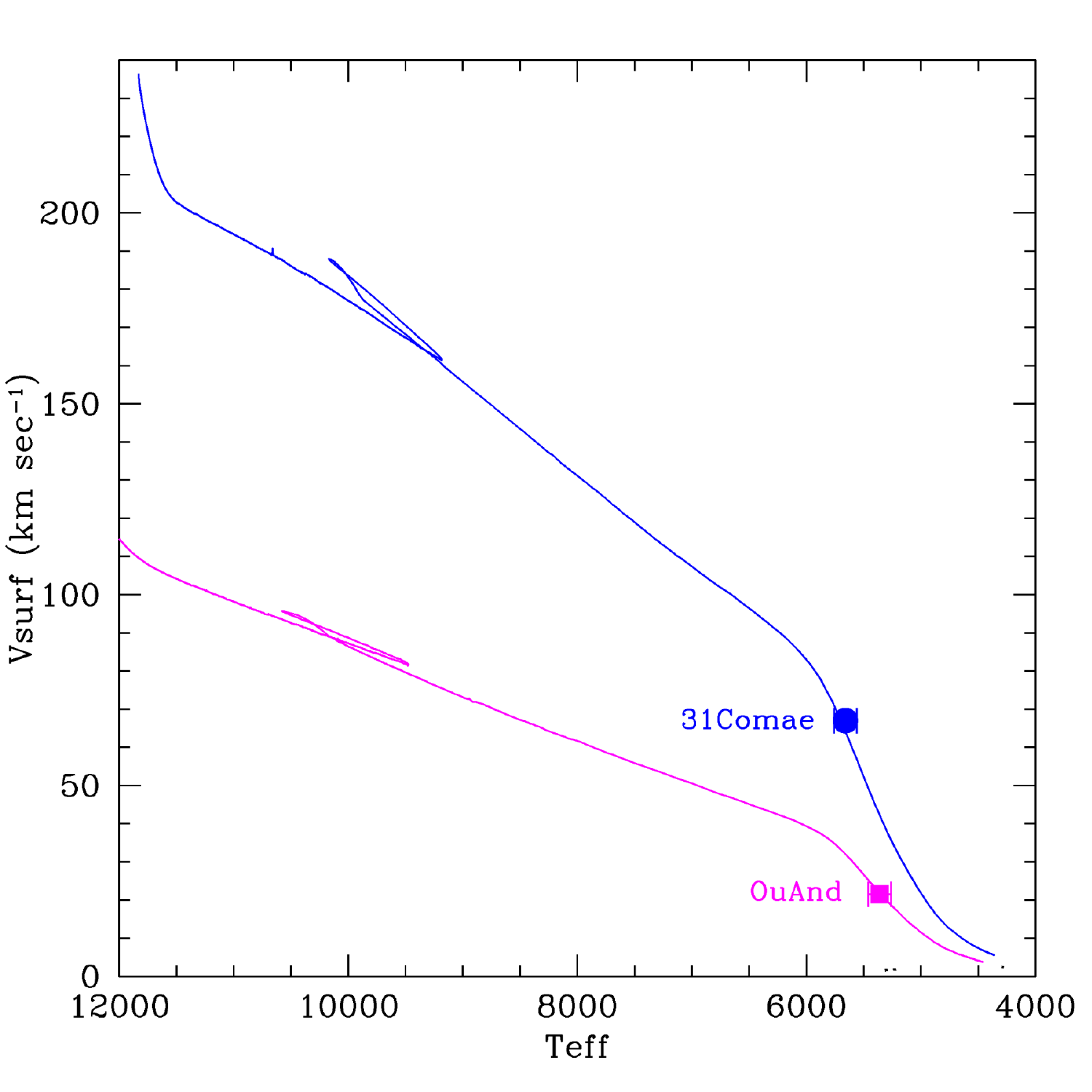}\\
\includegraphics[width=0.9\columnwidth]{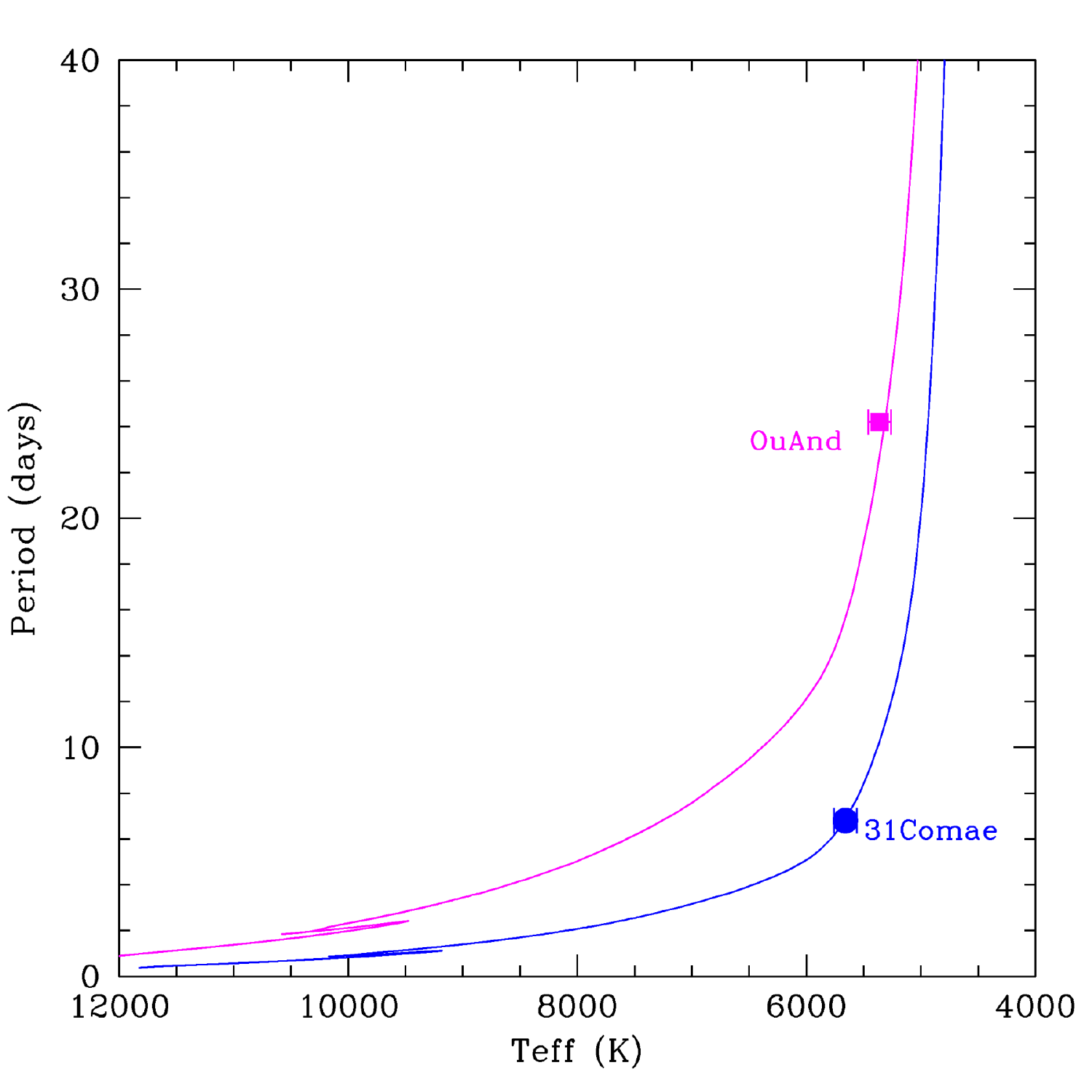}
\caption{Evolution of the surface rotation velocity (upper plot, in km sec$^{-1}$) and period (lower plot, in days) of the 2.7 and 2.85~M$_{\odot}$ dedicated models for 31~Com and OU~And (circle and square, respectively), compared with the observational data.}
\label{fig:modelsurfacerotationperiod}
\end{figure}

The positions of OU~And and 31~Com on the HR diagram are shown in Fig.~\ref{fig:evolution}.
The values we use for the stellar effective temperatures are shown in Table~\ref{tab:stars} and section \S~\ref{subsection:stellar_parameters}. Luminosity values are computed using the stellar parallaxes from the New Reduction Hipparcos catalogues by van Leeuwen (2007), the $V$ magnitudes from the 1997 Hipparcos catalogue \citep{hipp_97}, and applying the bolometric correction relation of \cite{flower_96}.

The evolution tracks shown in Fig.~\ref{fig:evolution} are for rotating models computed for solar metallicity with the code STAREVOL with the same input physics and assumptions as in \citet{Lagardeetal2012}. 
For the 2.5 and 3.0~M$_{\odot}$ models, the initial rotation velocity on the zero age main sequence V$_{\rm zams}$ corresponds to 30\% of the critical velocity at that stage \citep{Lagardeetal2014}.
For the present paper, we  computed models of 2.7 and 2.85~M$_{\odot}$ suited to 31~Com and OU~And, respectively, with different V$_{\rm zams}$ to infer the initial rotation velocities of OU~And and 31~Com.
No magnetic braking is applied, and the rotation velocity of the models evolves mainly owing to structural changes of the star (in particular changes of stellar radius). 

It is clear that both stars are in the Hertzsprung gap and have very similar masses of the order of $\sim$ 2.7 and 2.85$~M_{\odot}$ for 31~Com and OU~And, respectively, as already discussed in \cite{big_paper}. Initial rotation velocities of 235 and 131~km sec$^{-1}$, which correspond respectively to 53 and 30$\%$ of the critical velocity on the ZAMS, are inferred to reproduce the present day rotation velocities and periods of 31~Com and OU~And, respectively, as shown in Fig.~\ref{fig:modelsurfacerotationperiod}.

\begin{figure}[htb]
\centering
\includegraphics[width=0.90\columnwidth]{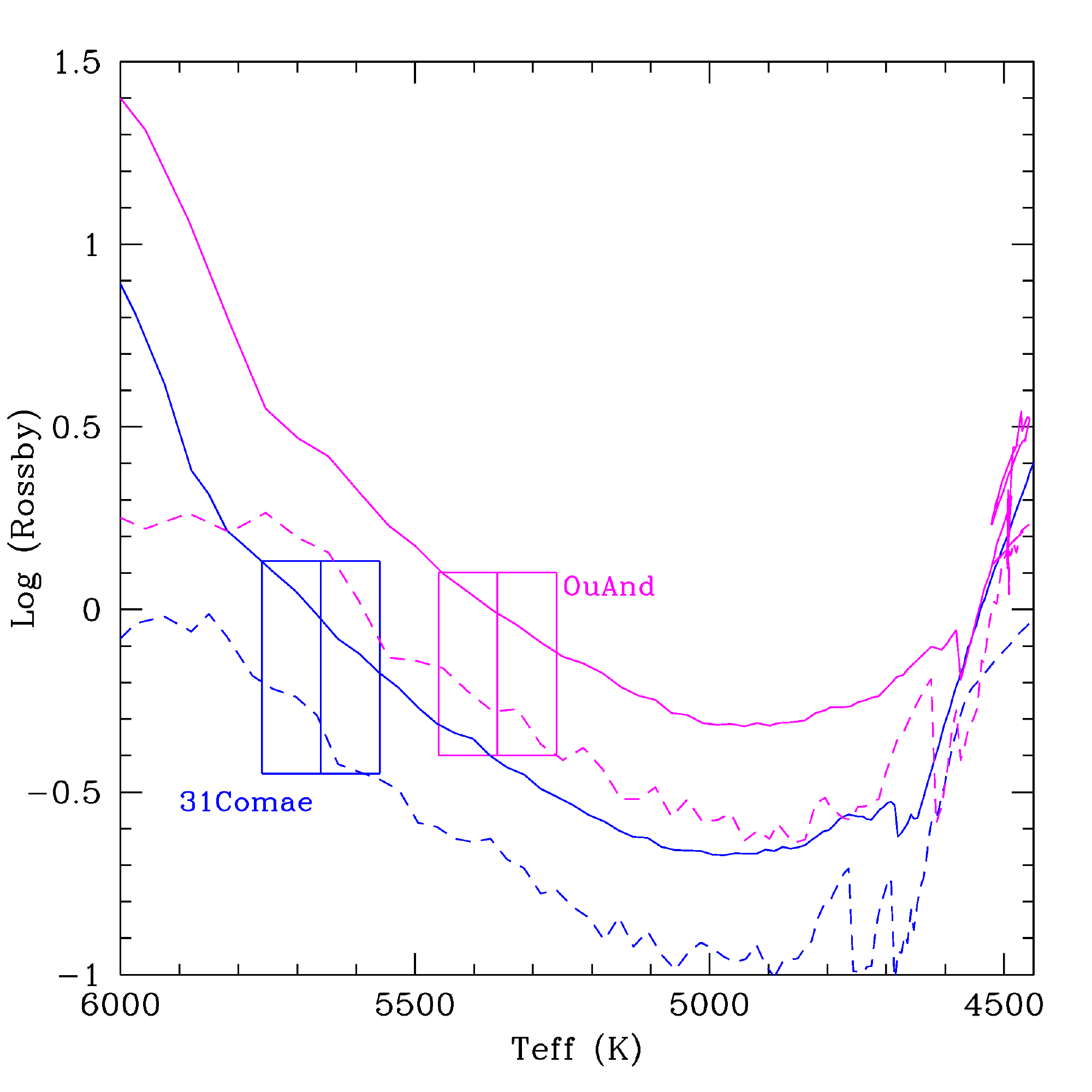}
\caption{Evolution of the theoretical Rossby number for the 2.7 and 2.85~M$_{\odot}$ dedicated models (blue and magenta respectively). 
The full and dotted lines correspond to Rossby numbers computed with the turnover timescale taken, respectively, at half a pressure scale height above the base of the convective envelope and at its maximum value within the convective envelope.
The rectangles correspond to the expected position of 31~Com and OU~And, taking into account uncertainties in their effective temperatures.}
\label{fig:tconv_rossby}
\end{figure}

\section{Discussion: Origin of the magnetic field and activity of OU~And and 31~Com}

Figure~\ref{fig:tconv_rossby} shows the evolution of the theoretical 
Rossby number for the 2.7 and 2.85~M$_{\odot}$ models that fit the 
present-day rotation of 31~Comae and OU~And (\S~\ref{subsection:evolution}). R$_O$ is defined as the ratio between 
the rotational period and the convective turnover timescale, P$_{rot} / 
\tau_c$, and these quantities come directly from the stellar evolution 
computations.
For each stellar mass we show  cases when the Rossby number is 
computed using the turnover timescale at half a pressure scale height 
above the base of the convective envelope and at its maximum value 
within the convective envelope (full and dotted lines, respectively).

We see that the theoretical Rossby number decreases towards lower effective temperatures, 
owing to the fast increase of the convective turnover timescale as the 
convective envelope deepens in mass during the first dredge-up in the 
so-called magnetic strip (Charbonnel et al., in prep.).

 Figure~\ref{fig:tconv_rossby} shows that both stars have rather small Rossby number reaching about 0.5 at maximum value for $\tau_{conv}$. This suggests that a dynamo-driven magnetic field could  therefore be at work, which would be the origin of the large activity of these stars. However, these stars have significantly different rotation periods and magnetic field strengths. Better understanding of the magnetic activity of both stars gives the comparison of the magnetic field strength with the semi-emperical R$_O$ defined as the ratio between the observed $P_{rot}$ and the maximum value of the convective turnover timescale within the convective envelope. For OU~And, the semi-emperical R$_O$ is $0.68$ \citep{big_paper}, and it fits with our theoretical predictions. For 31~Comae it is $0.12$, \citep{big_paper}, and it is about two times greater than our theoretical predictions. 
The faster rotator, 31~Com ($P_{rot}$ = 6.8~d) has magnetic field properties that are consistent with an $\alpha - \omega$ dynamo-driven magnetic field \citep{big_paper} and the magnetic map obtained in Section \S~\ref{subsection:MM31Com} is consistent with this hypothesis.

\begin{figure}[htb]
\centering

\includegraphics[width=0.95\columnwidth]{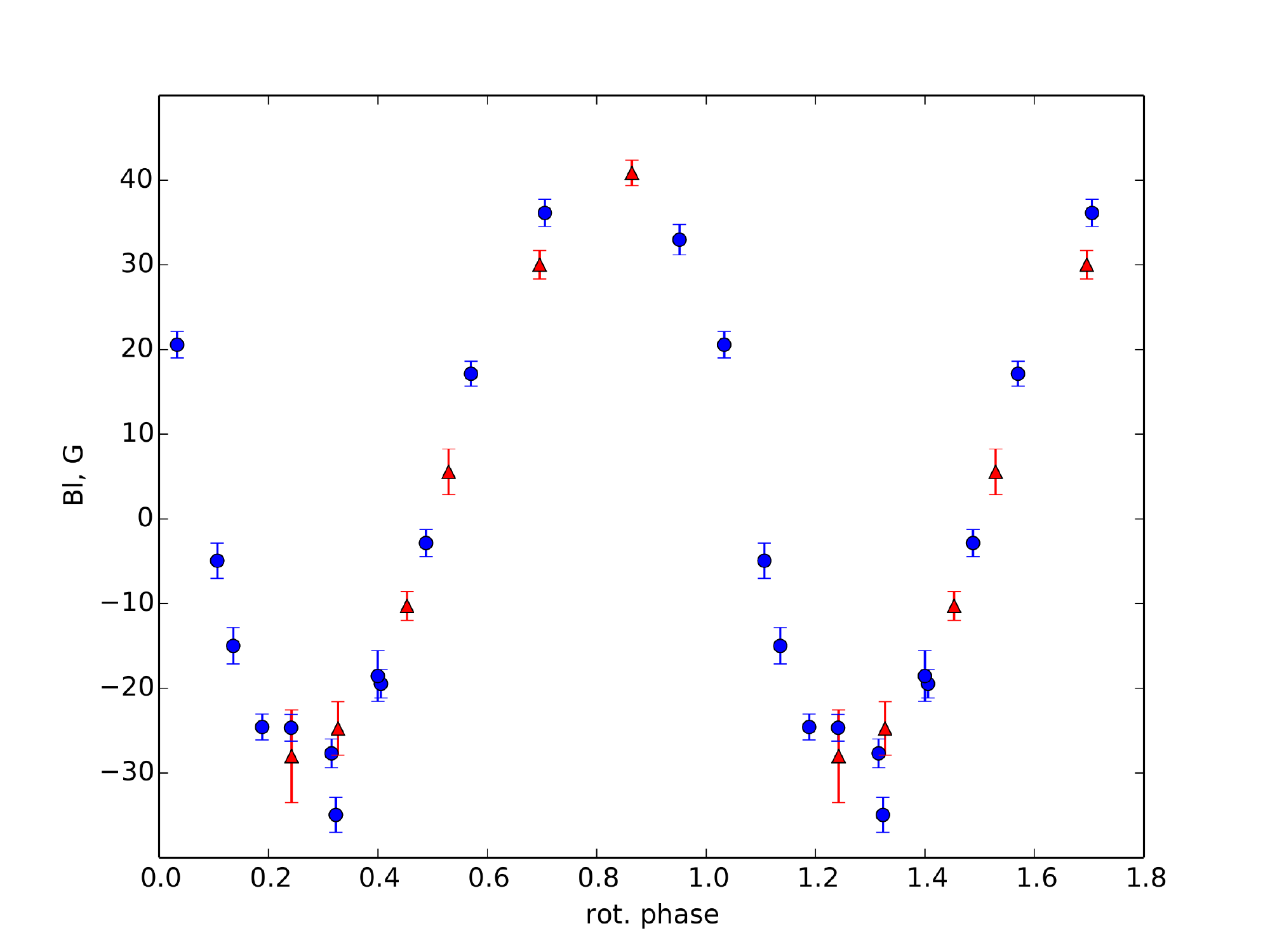}

\caption{$B_l$ of OU~And in 2013 (circles) and in 2008 shifted with 0.5 cycle (triangles).}

\label{fig:Bl_shift}
\end{figure}

On the other hand, though OU~And appears with a rather fast rotation period of 24.2~days, it is a deviating object on several plots presented in the study of \cite{big_paper}, which led these authors to suppose that OU~And is the probable descendant of a magnetic Ap star on the main sequence. Our ZDI study supports arguments to this hypothesis: its magnetic topology is mainly poloidal (82\% of the magnetic energy in 2008) and the poloidal component is dominated by a dipole; furthermore, this topology appears to be quite stable. Figure~\ref{fig:2profiles} (right panel) illustrates the stability of the magnetic field from one rotation to the next one. A comparison of the maps obtained in 2008 (lower panel of Fig.~\ref{fig:ouand08_a}) and in 2013 (left panel of Fig.~\ref{fig:maps}) and the magnetic field parameters presented in Table~\ref{tab:dynamo} show that the magnetic topology is remarkably similar five years apart. Figure \ref{fig:Bl_shift} also supports this statement and demonstrates that the 2008 $B_l$ 
data, shifted with by half a cycle, overlap the rotational modulation of $B_l$ in 2013. This might be an indication that the 0.5-cycle shift could be attributed to the precision of the period determination. However Table~\ref{tab:dynamo} shows that when the magnetic field strength increases between 2008 and 2013, the percentage strength of the poloidal component decreases from 82\% to 64\%, which corresponds to a correlated increase in the toroidal component. At the same time, the dipole contribution of the poloidal component remains the same. This indicates that the magnetic strength variation is mainly due to the toroidal component (probably of rotational dynamo origin), when the dipolar component (presumed to be of fossil origin) - remains about constant. This is consistent with our analysis that magnetic field of OU~And is due to the interplay between convection and the remnant of the fossil field of an Ap~star. Following the approach of \cite{stepien_93}, with the assumption of magnetic flux 
conservation and a radius increase with stellar evolution to the red-giant branch (RGB), we can infer the dipole strength of a possible Ap~star progenitor of OU~And. Having a radius on the Main Sequence (MS) of 2.2~$R_{\odot}$, obtained by the proposed evolutionary model and the mean and maximum poloidal component of the magnetic field strength in 2013 (Table~\ref{tab:dynamo}), we 
find the initial values for magnetic dipole strength 
on the main sequence of the HRD of about 800 to 4600~G, respectively, consistent with  values for Ap~stars. The predicted rotation of OU~And on the MS, with $v\sin(i)$ of about 131~km sec$^{-1}$, is faster than  the mean for Ap stars but is still consistent with the Ap star status. Very few Ap stars are known to have fast rotation ($v\sin(i)$ over 100 km/s), but this type of small sample does exist, according to \cite{wraight_2015}.

Direct comparison of the properties of OU~And and 31~Com emphasizes the differences that were listed just above. Both stars have about the same mass, about the same evolutionary state, and the same order of high level X-ray activity. Though 31~Com rotates three times faster than OU~And, the magnetic strength of OU~And is twice that of 31~Com. Figure~\ref{fig:stokesv} and Figure~\ref{fig:maps} clearly show the difference in magnetic topology: a large scale dominant structure for OU~And, a smaller scale dominant structure for 31~Com. While the magnetic map of 31~Com can be compared to that of 390~Aur \citep{renada_v390}, the outstandingly strong large scale magnetic field of OU~And is reminiscent of the properties observed on EK~Eri, which is the archetype of the descendants of magnetic Ap~stars on the main sequence \citep{stepien_93,strass_99,auriere_2008, auriere_2012}.
 We also need to mention that the dynamo models, suitable for some M~dwarfs and capable of producing fields like the one for OU~And, are not effective in this case, because M~dwarfs have huge convective zones or are fully convective for spectral types that are cooler than M3 \citep{morin_2010}. On the other hand, in the case of intermediate mass stars in the Hertzsprung gap, the convective envelope is just beginning to develop and a possible dynamo mechanism starts to work. If a toroidal field at the surface of the star is detected, this is a strong indication of a distributed dynamo \citep{donati_2003}. A dynamo-generated magnetic field at this evolutionary stage could be produced in a fast rotator by a dynamo mechanism (the case of 31~Comae), as well as from the Ap~ star fossil magnetic field interacting with relatively fast remnant rotation. Without a remnant Ap~magnetic field, OU~And with its rotational rate should be less active and with a weaker magnetic field than 31~Comae.
 
\section{Conclusions}
It is well known that intermediate mass stars on the MS do not have a convective envelope as well as nor do they have dynamo-driven surface magnetic fields and winds that carry away angular momentum and slow down the rotation. A recently published paper \citep{stello_2016} demonstrates that most MS intermediate mass stars host a strong dynamo in their core,  the product of this dynamo  being confined deep in the star and, consequently, does not show  at surface level. 

On the other hand, we have studied the case of two evolved intermediate stars with magnetic activity and, for the first time, performed  the ZDI magnetic mapping of OU~And and 31~Com. Thanks to  dedicated evolutionary models, including rotation \citep{charb_2010}, we can precise the evolutionary status of these stars in the Hertzsprung gap and the history of their rotation, which has not been affected by magnetic braking, neither on the main sequence nor after, hence magnetic braking plays no role until the star develops a cool star structure.

Observational properties of red giants, i.e., fast rotators with dynamo magnetic fields are: a Rossby number smaller than unity, peaks and bumps in the Stokes~I profile, a complex structure in the magnetic map and a significant toroidal component of the MF. Depending on the thickness of the convective envelope and rotational rate in some stars, we also can observe an azimuthal belt in the magnetic map \citep[e.g., V390 Aur,][]{renada_v390}. There is also a dependence of the magnetic field strength on rotation \citep{renada_2013, big_paper}, which is clear evidence of an $\alpha$-$\omega$ dynamo operation.

Both the individual analysis of the magnetic fied and activity of each star and a direct comparison of their properties show that they illustrate the evolution of a fast rotator on the one hand (31~Com), and of a probable Ap star descendant (OU~And), on the other. The origin of the magnetic field at the surface of 31~Com is suspected to be dynamo-driven. The dipolar magnetic field component of OU~And could be the remnant of of an Ap~star on the main sequence, now interacting with convection. The toroidal magnetic field component of OU~And may be due to a dynamo as a result of the fast rotation of the star. The two stars are observed as being still at the beginning of their evolution off the main sequence: the convective envelope of 31~Com is still rather shallow and the magnetic dipole strength from the Ap star progenitor of OU~And is diluted by about 18 times. In its present evolutionary state, the magnetic field of OU~And is significantly stronger than that of 31~Com, as it is for for the \ion{Ca}{ii}~H\&K 
and H$\alpha$ lines, too. The two stars are more similar for their $L_{x}$ and the  strength of other activity indicators. In the sample studied by \cite{big_paper}, only one star of similar mass (3~$M_{\odot}$), $\iota$~Cap, has such a high rotaional rate with measured $P_{rot}$ (68~d). This is located at the base of the RGB, and it does not deviate from the mean relation between rotation and magnetic field strength observed by Auri\`ere et al. (2015): its magnetic field is likely to be dynamo-driven and may illustrate the changes of rotational and magnetic field properties  expected for 31~Com when reaching the same evolutionary status. The $|B_l|_{max}$ of $\iota$~Cap is of 8.3~G, which is about the same as for 31~Com in its present state in the Hertzsprung gap. This may indicate that, at the evolutionary state of $\iota$~Cap (at the base of the RGB) when the convective envelope deepens and the convective turnover time increases but the rotation rate decreases (because of magnetic 
braking and increased radius) the magnetic field strength produced by the dynamo of 31~Comae will remain of the same magnitude. However, at the same evolutionary stage, using the models presented in Auri\`ere et al. (2015) and the hypothesis for magnetic flux conservation, the magnetic strength of the dipole of OU~And  only drops by a factor of 2, and is still significantly stronger than the magnetic strength of 31~Com. These extrapolations suggest that the Ap-star descendants, even with moderate magnetic dipoles like OU~And, will remain outstanding objects with respect to the period/magnetic strength relation up until they begin  ascend the RGB.

\begin{acknowledgements}
We are thankful to the TBL team for providing service observations with Narval spectropolarimeter. 
Observations in 2013 were funded under the project BG051PO001-3.3.06-0047 financed by the EU, ESF, 
and Republic of Bulgaria. Observations in 2008 were funded under an OPTICON project. This work was also supported by the Bulgarian NSF contracts DRILA 01/03 and DMU 03-87. N.A.D. thanks Saint Petersburg State University, Russia, for research grant 6.38.18.2014 and FAPERJ, Rio de Janeiro -Brazil, for visiting researcher grant E-26/200.128/2015. The authors gratefully acknowledge the constructive comments and input offered by the referee.
\end{acknowledgements}

\bibliography{aaouand.bib}
%-------------------------------------------------------------------

\begin{appendix}
\section{Determining the  atmospheric parameters of OU~And}
\begin{figure}[htb]
\centering
\includegraphics[width=0.98\columnwidth]{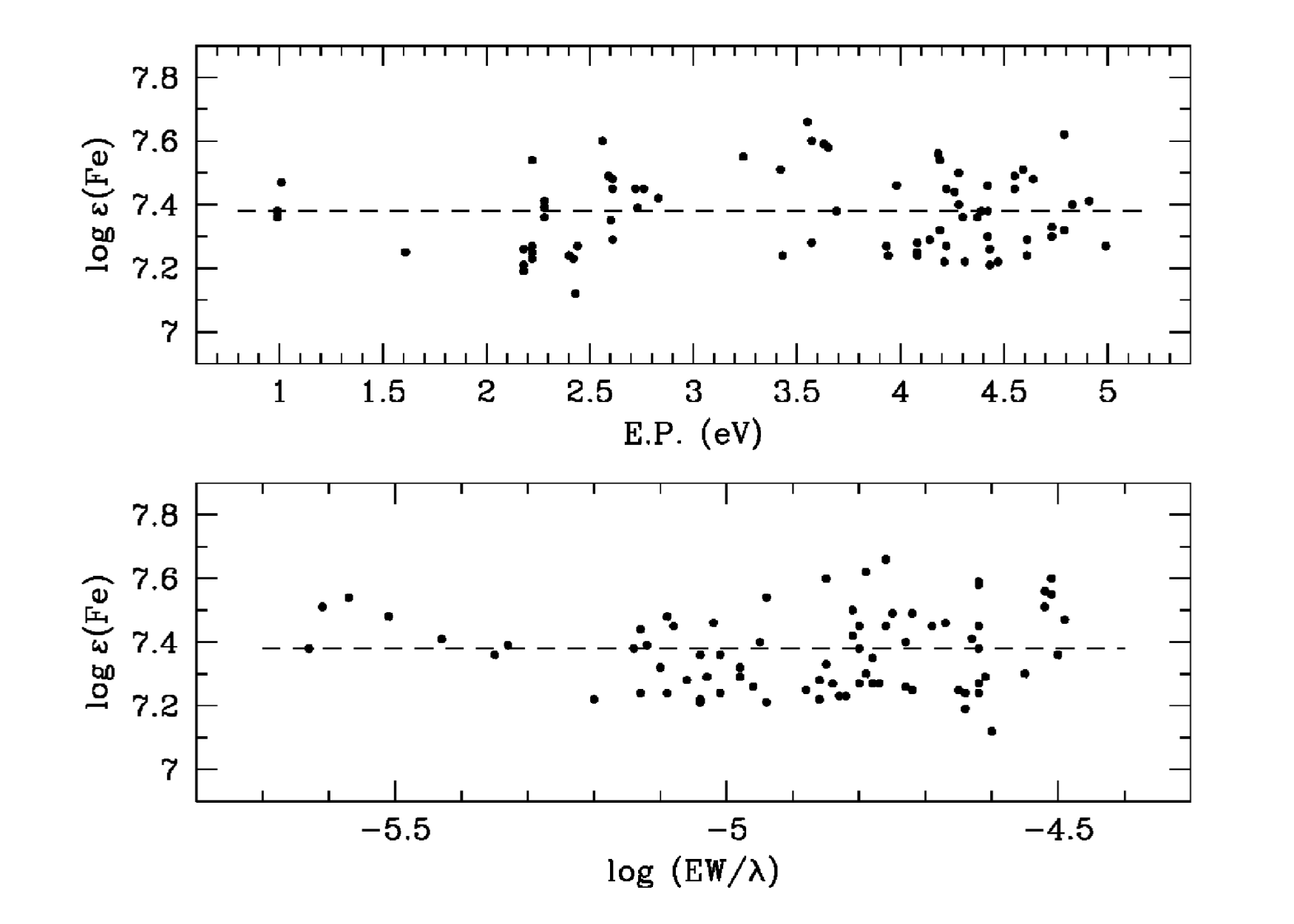}
\label{fig:FeEWs}
\caption{Abundance of \ion{Fe}{i} lines versus excitation potential.
The correct temperature is indicated by a zero slope in this plot.
Bottom: Abundance of \ion{Fe}{i} lines versus showing determination of the microturbulent velocity.}
\end{figure}
\newpage

\onecolumn
\begin{center}
\begin{longtable}{ccccc}
\caption[]{\ion{Fe}{i} and \ion{Fe}{ii} lines used for atmospheric
parameter determination. The wavelengths are given in the first column,
the lower excitation potentials of the transitions ($\chi$, eV) and
the oscillation strengths ($\log gf$) are given in the second and third
columns. The last column provides the measured equivalent widths (EW, m\AA) in the spectrum of OU~And.}\\

\hline
\hline
\multicolumn{1}{c}{Element} & 
\multicolumn{1}{c}{$\lambda$} &
\multicolumn{1}{c}{$\chi$ (eV)} &
\multicolumn{1}{c}{$\log gf$} &
\multicolumn{1}{c}{EW (m\AA)} \\
\hline
 \endfirsthead
 
 \hline
 \hline
\multicolumn{1}{c}{Element} & 
\multicolumn{1}{c}{$\lambda$} &
\multicolumn{1}{c}{$\chi$ (eV)} &
\multicolumn{1}{c}{$\log gf$} &
\multicolumn{1}{c}{EW (m\AA)} \\
 \hline
 \endhead

 \hline
 \endfoot
 \hline\hline
 \endlastfoot
 
\ion{Fe}{i} &  5159.058 &  4.283     & 2.239e-01&       96.7 \\
 &  5162.273 &  4.178& 0.120e+01&      156.9 \\
 & 5198.711  &  2.223& 7.244e-03&      125.4 \\
 & 5242.491  &  3.634& 1.072e-01&      125.2 \\
 & 5288.525  &  3.694& 3.090e-02&       83.6 \\
 & 5307.361  &  1.608& 1.072e-03&      118.5 \\
 & 5315.051  &  4.371& 3.981e-02&       48.4 \\
 & 5321.108  &  4.434& 6.457e-02&       48.2 \\
 & 5367.467  &  4.415& 0.275e+01&      150.5 \\
 & 5373.709  &  4.473& 1.950e-01&       74.5 \\
 & 5389.479  &  4.415& 5.623e-01&      116.3 \\
 & 5393.168  &  3.241& 1.906e-01&      165.4 \\
 & 5441.339  &  4.312& 2.630e-02&       34.6 \\
 & 5445.042  &  4.386& 0.110e+01&      131.5 \\
 & 5497.516  &  1.011& 1.445e-03&      176.5 \\
 & 5506.779  &  0.990& 1.585e-03&      173.5 \\
 & 5522.447  &  4.209& 3.981e-02&       50.8 \\
 & 5531.984  &  4.913& 3.467e-02&       20.7 \\
 & 5532.747  &  3.573& 0.100e-01&       77.3 \\
 & 5554.895  &  4.548& 4.169e-01&      104.9 \\
 & 5560.212  &  4.434& 9.120e-02&       61.3 \\
 & 5567.391  &  2.608& 2.754e-03&       95.7 \\
 & 5569.618  &  3.417& 3.236e-01&      169.1 \\
 & 5576.089  &  3.430& 1.413e-01&      127.8 \\
 & 5584.765  &  3.573& 6.761e-03&       48.7 \\
 & 5633.947  &  4.991& 7.586e-01&       82.0 \\
 & 5635.823  &  4.256& 1.820e-02&       41.5 \\
 & 5638.262  &  4.220& 1.905e-01&       92.6 \\
 & 5691.497  &  4.301& 4.266e-02&       55.6 \\
 & 5717.833  &  4.284& 1.050e-01&       87.9 \\
 & 5806.725  &  4.607& 1.259e-01&       61.1 \\
 & 5934.655  &  3.928& 9.550e-02&       94.4 \\
 & 6016.660  &  3.546& 2.138e-02&      103.4 \\
 & 6024.058  &  4.548& 8.710e-01&      123.4 \\
 & 6027.051  &  4.076& 8.128e-02&       82.9 \\
 & 6056.005  &  4.733& 3.981e-01&       86.1\\
 & 6065.482  &  2.608& 2.951e-02&      147.4\\
 & 6082.711  &  2.223& 2.630e-04&       69.8\\
 & 6096.665  &  3.984& 1.660e-02&       58.8\\
 & 6151.618  &  2.176& 5.129e-04&       70.8\\
 & 6157.728  &  4.076& 7.762e-02&       81.0\\
 & 6165.360  &  4.142& 3.388e-02&       57.1\\
 & 6170.507  &  4.795& 4.168e-01&      100.7\\
 & 6173.336  &  2.223& 1.318e-03&       94.4\\
 & 6187.990  &  3.943& 2.692e-02&       61.0\\
 & 6200.313  &  2.605& 3.631e-03&      101.8\\
 & 6213.430  &  2.223& 3.311e-03&      117.7\\
 & 6230.723  &  2.559& 5.248e-02&      192.7\\
 & 6252.555  &  2.403& 1.906e-02&      148.9\\
 & 6265.130  &  2.180& 2.818e-03&      117.5\\
 & 6322.686  &  2.588& 3.715e-03&      112.0\\
 & 6380.743  &  4.186& 4.786e-02&       66.6\\
 & 6392.539  &  2.280& 9.333e-05&       30.2\\
 & 6393.601  &  2.433& 3.715e-02&      159.1\\
 & 6411.649  &  3.653& 2.188e-01&      155.0\\
 & 6419.950  &  4.733& 8.128e-01&      104.1\\
 & 6421.351  &  2.279& 9.772e-03&      151.9\\
 \ion{Fe}{i}& 6430.846  &  2.176& 9.772e-03&      145.8\\
 & 6436.407  &  4.190& 3.467e-03&       17.5\\
 & 6469.193  &  4.835& 2.399e-01&       72.4\\
 & 6518.367  &  2.830& 5.012e-03&      102.0\\
 & 5417.033  &  4.415& 2.951e-02&       39.5 \\
 & 6551.678  &  0.990& 1.621e-06&       15.5\\
 & 6591.313  &  4.590& 8.511e-03&       16.1\\
 & 6592.914  &  2.723& 3.388e-02&      159.6\\
 & 6593.871  &  2.437& 3.802e-03&      111.7\\
 & 6597.561  &  4.795& 1.202e-01&       52.5\\
 & 6608.026  &  2.280& 9.333e-05&       29.5\\
 & 6646.932  &  2.610& 1.023e-04&       20.6\\
 & 6703.567  &  2.760& 6.918e-04&       55.9\\
 & 6750.153  &  2.424& 2.399e-03&       99.7\\
 & 6752.707  &  4.638& 6.310e-02&       55.4\\
 & 6806.845  &  2.730& 6.166e-04&       51.5\\
 & 6810.263  &  4.607& 1.023e-01&       55.6\\
 & 7130.922  &  4.217& 1.995e-01&      112.5\\
 & 7132.986  &  4.076& 2.455e-02&       53.3\\
 \hline
 \ion{Fe}{ii} & 5132.657  &  2.807& 1.000e-04&       51.6\\
 & 6084.099  &  3.199& 1.585e-04&       45.0\\
 & 6149.246  &  3.889& 1.905e-03&       67.0\\
 & 6247.545  &  3.891& 4.571e-03&       89.8\\
 & 6416.921  &  3.891& 2.088e-03&       69.0\\
 & 6432.682  &  2.891& 2.630e-04&       75.4\\
\hline
%\end{tabular}
\end{longtable}
\end{center}

\end{appendix}

\end{document}